\documentclass[letterpaper,10pt]{article}
\usepackage[top=0.85in,left=1in,footskip=0.75in,marginparwidth=1in]{geometry}

\usepackage{array}
\usepackage{changepage}
\usepackage{defs}
\usepackage{psfig}
\usepackage{graphicx}
\usepackage{amsmath}
\usepackage{amssymb}
\usepackage{natbib}
\usepackage{enumitem}
\usepackage{float}
\usepackage{caption}
\usepackage{epstopdf}
\usepackage{appendix}
\usepackage{placeins}
\usepackage{url} 
\usepackage{lipsum}
\usepackage{multirow}
\usepackage{gensymb}
\usepackage{siunitx}

\begin{document}

\begin{center}
{\huge
\textbf{Automated Cross-identifying Radio to \\
Infra-red Surveys Using the \\
LRPY Algorithm: A Case Study.}
}
\end{center}

S.D. Weston\textsuperscript{1,2},
N. Seymour\textsuperscript{3}, 
S. Gulyaev\textsuperscript{1},  
R.P. Norris\textsuperscript{2,4}, 
J. Banfield\textsuperscript{6}, 
M. Vaccari\textsuperscript{7,8}, 
A.M. Hopkins\textsuperscript{5}, 
T.M.O. Franzen\textsuperscript{3,2}
\\\\
\bf{1} Institute for Radio Astronomy \& Space Research, AUT University, Auckland, New Zealand\\
\bf{2} CSIRO, P.O. Box 76, Epping, NSW 1710, Australia\\
\bf{3} International Centre for Radio Astronomy Research, Curtin University, Perth, Australia\\
\bf{4} Western Sydney University, Penrith, NSW, Australia\\
\bf{5} Australian Astronomical Observatory, PO Box 915, North Ryde, NSW 1670, Australia\\
\bf{6} Research School of Astronomy and Astrophysics, Australian National University, Canberra ACT, 2611, Australia\\
\bf{7} Department of Physics and Astronomy, University of the Western Cape, Robert Sobukwe Road, 7535 Bellville, Cape Town, South Africa\\
\bf{8} INAF - Istituto di Radioastronomia, via Gobetti 101, 40129 Bologna, Italy
\\\\
* stuart.weston@aut.ac.nz

\vspace*{1\baselineskip}
\begin{center}
\Large\today
\end{center}

\raggedbottom

\section*{Abstract}
Cross-identifying complex radio sources with optical or infra red (IR) counterparts in surveys such as the Australia Telescope Large Area Survey (ATLAS) has traditionally been performed manually. However, with new surveys from the Australian Square Kilometre Array Pathfinder (ASKAP) detecting many tens of million of radio sources such an approach is no longer feasible. This paper presents new software (LRPY - Likelihood Ratio in PYthon) to automate the process of cross-identifying radio sources with catalogues at other wavelengths. LRPY implements the Likelihood Ratio (LR) technique with a modification to account for two galaxies contributing to a sole measured radio component. We demonstrate LRPY by applying it to ATLAS DR3 and a {\it Spitzer}-based multi-wavelength fusion catalogue, identifying 3,848 matched sources via our LR-based selection criteria. A subset of 1987 sources have flux density values for all IRAC bands which allow us to use criteria to distinguish between active galactic nuclei (AGN) and star-forming galaxies (SFG). We find that 936 radio sources ($\approx47\,\%$) meet both of the Lacy and Stern AGN selection criteria. Of the matched sources, 295 have spectroscopic redshifts and we examine the radio to IR flux ratio vs redshift, proposing an AGN selection criterion below the Elvis radio-loud (RL) AGN limit for this dataset. Taking the union of all three AGN selection criteria we identify 956 as AGN ($\approx48\,\%$). From this dataset, we find a decreasing fraction of AGN with lower radio flux densities consistent with other results in the literature. 

\clearpage

\section{INTRODUCTION}



By obtaining large datasets at different wavelengths, which are sensitive to different galaxy properties, one can separate the different influences on the formation and evolution of galaxies. One key problem in combining these multi-wavelength surveys is determining which sources, observed at different wavelengths, are truly associated with one another and which are unrelated. We are now entering a new epoch of radio astronomy with even greater galaxy surveys such as EMU \citep[the Evolutionary Map of the Universe,][]{Norris:2012} and the MIGHTEE \citep[MeerKAT International GigaHertz Tiered Extragalactic Exploration survey,][]{Heyden:2010} 
using the Square Kilometre Array precursors ASKAP and MeerKAT, which will increase the number of faint (sub-mJy) radio sources by orders of magnitude. All of these faint radio sources will need to 
be cross-matched with other wavelength datasets in order to tackle numerous science questions, in particular the star formation history of the Universe, a key goal for the SKA \citep{Prandoni:2011}. Where surveys have similar wavelengths, resolutions and sensitivities this matching of sources is generally straightforward and unambiguous (e.g. by simple nearest-neighbour matching). However, where the surveys are dissimilar, with very different resolutions and/or sensitivities, such as between radio and infrared, a nearestneighbour approach becomes unreliable. Furthermore, radio sources may have complex structure which is very different to the appearance at optical and near-infrared wavelengths.

There are methods which overcome the problem of cross-matching different wavelengths, such as the Likelihood Ratio (LR) method first proposed by \citet{Richter:1975}, the Poisson Probability (PP) method \citet{Downes:86} and a Bayesian method \citet{Fan:2015}. The LR method not only uses positional information but also flux and source density. In the rare complex cases, beyond the capabilities of computational algorithms, human pattern recognition can be used via citizen science projects such as Radio Galaxy Zoo \citep{Banfield:2016}. Despite having many participants, such methods are time consuming with each cross-match requiring ten or more classifications. Hence, with potentially 70 million sources to be detected by EMU, it is desireable to maximise the fraction which can be cross-matched in an automated way.

In this paper we present an implementation of the LR method within an algorithm for cross-identifying sources between a radio catalogue and an infrared (IR) catalogue. The Likelihood Ratio in PYthon code (LRPY) developed for this paper is also realised to the public. 

The paper is structured as follows. In Section 2 we present the ATLAS and Fusion surveys. In Section 3 we present our implementation of the LR technique. In Section 4 we review the selection criteria for possible true matches and to identify multiple components in the infrared domain for one radio source. In Section 5 we take the cross-matches between radio and infrared and examine some of their colour-colour properties and radio to infrared flux as a function of redshift. We present an additional catalogue to ATLAS DR3 in Section 6 by including our matches with the Spitzer Data Fusion catalogue with the corresponding LR and Reliability values.


\section[]{DATA}

This work and analysis concentrates on the ATLAS DR3 catalogue \citep{Franzen:2015} and the {\it Spitzer} Data Fusion catalogue \citep{Vaccari2010,Vaccari2015}\footnote{\url{http://mattiavaccari.net/df/}}.

 
\subsection{ATLAS DR3 Radio Catalogue} \label{atlas_dr3_radio_catalogue}

The ATLAS survey was conducted with the Australia Telescope Compact Array (ATCA) between April 2002 and June 2010 and covers 1.3--1.8\,GHz, over an area coinciding with the {\it Chandra} Deep Field South (CDFS) and the European Large Area ISO Survey - South 1 \citep[ELAIS-S1][]{RowanRobinson:99}. 
These areas also coincide with the {\it Spitzer} Wide-area Infrared Extragalactic survey (SWIRE, \citealt{Lonsdale2003}), thus providing the multi-wavelength data required for identification of the radio sources and further analysis (e.g. \citealt{Mao2012}). ATLAS Data Release 1 (DR1) was presented in \citet{Norris:07} and \citet{Middelberg:2007} and DR2 in \citet{Hales:14}.

The ATLAS DR3 component source catalogue (hereafter referred to as the `ATLAS' catalogue) presented in \citet{Franzen:2015} found a total of 5191 radio source components over $5\sigma$ for both fields. There are 3078 source components in CDFS above $70\,\mu$Jy $beam^{-1}$ and 2113 source components above $85\,\mu$Jy $beam^{-1}$ in ELAIS-S1. The restoring beam parameters used in our subsequent analysis for the two ATLAS fields are given in Table \ref{table:radio_beam}.
\citet{Middelberg:2007} identified a positional offset between the ATLAS DR1 and SWIRE catalogues in ELAIS-S1. A systematic offset of mean value $0.08 \pm 0.03\,''$  in right ascension and $0.06 \pm 0.03\,''$ in declination was found. We find an identical offset between the ATLAS DR3 and Fusion catalogues in our analysis and apply this to the ATLAS catalogue. 

\begin{table}
 \centering
  \caption{Restoring Beam for each ATLAS radio image }
  \begin{tabular}{@{}lrrr@{}}
  \hline
  \hline
   Field & $\Phi_{\rm Maj}$  & $\Phi_{\rm Min}$    & Position Angle   \\
           & (arcsec)      & (arcsec)       & (degrees) \\
  \hline
    CDFS    & 16.8  & 6.9 & 1.0   \\
    ELAIS-S1& 12.2  & 7.6 & -11.0 \\
  \hline
  \hline
\end{tabular}
\label{table:radio_beam}
\end{table}



\subsection{UV to Mid-IR Fusion Catalogue}

The Spitzer Data Fusion catalogue is a multi-wavelength far-UV to mid/far-IR catalogue of {\it Spitzer} selected sources, hereafter referred to as the `Fusion' catalogue \citep{Vaccari2012,Vaccari2015}, which covers the CDFS and ELAIS-S1 fields. This catalogue is based on detections at $3.6\,\mu m$ with the IRAC instrument \citep{Fazio:04} on board the {\it Spitzer Space Telescope} \citep{Werner:04}, down to a flux
density of $4.6\,\mu$Jy in the CDFS field and $4.8\,\mu$Jy in the ELAIS-S1 field. There are 391,518 {\it Spitzer}/IRAC sources in ELAIS-S1 and 462,638 in CDFS. 
The cross-identification performed in this analysis
makes use of the IRAC $3.6\,\mu m$ flux density. We note that this catalogue contains very few photometric and spectroscopic redshifts pertaining to radio sources, but the Herschel Extragalactic Legacy Project (HELP) \citet{Vaccari2016} is in the process of putting together multi-wavelength data, compute photometric redshifts and physical parameters for sources in ATLAS (and ASKAP/EMU Early Science) fields.




\subsection{OzDES data}
\label{ozdes}
OzDES \citep{Yuan:2015} a multi-year, 100\,night spectroscopic survey of the Dark Energy Survey \citep[DES][]{DES:16} deep supernova fields with the 4m Anglo-Australian Telescope. While the primary goal is to obtain spectra of supernovae detected by DES and their host galaxies, the design of the survey allows other projects to target sources in these fields. Five of the DES deep fields overlap with CDFS and ELAIS-S1, hence OzDES is able to provide spectra of many radio sources. In this work we use the 2016-02-25 version of the Global Redshift Catalogue which includes OzDES and literature spectra in the DES deep fields. 


\noindent
\begin{figure*}
\centering
     \includegraphics[width=0.95\columnwidth]{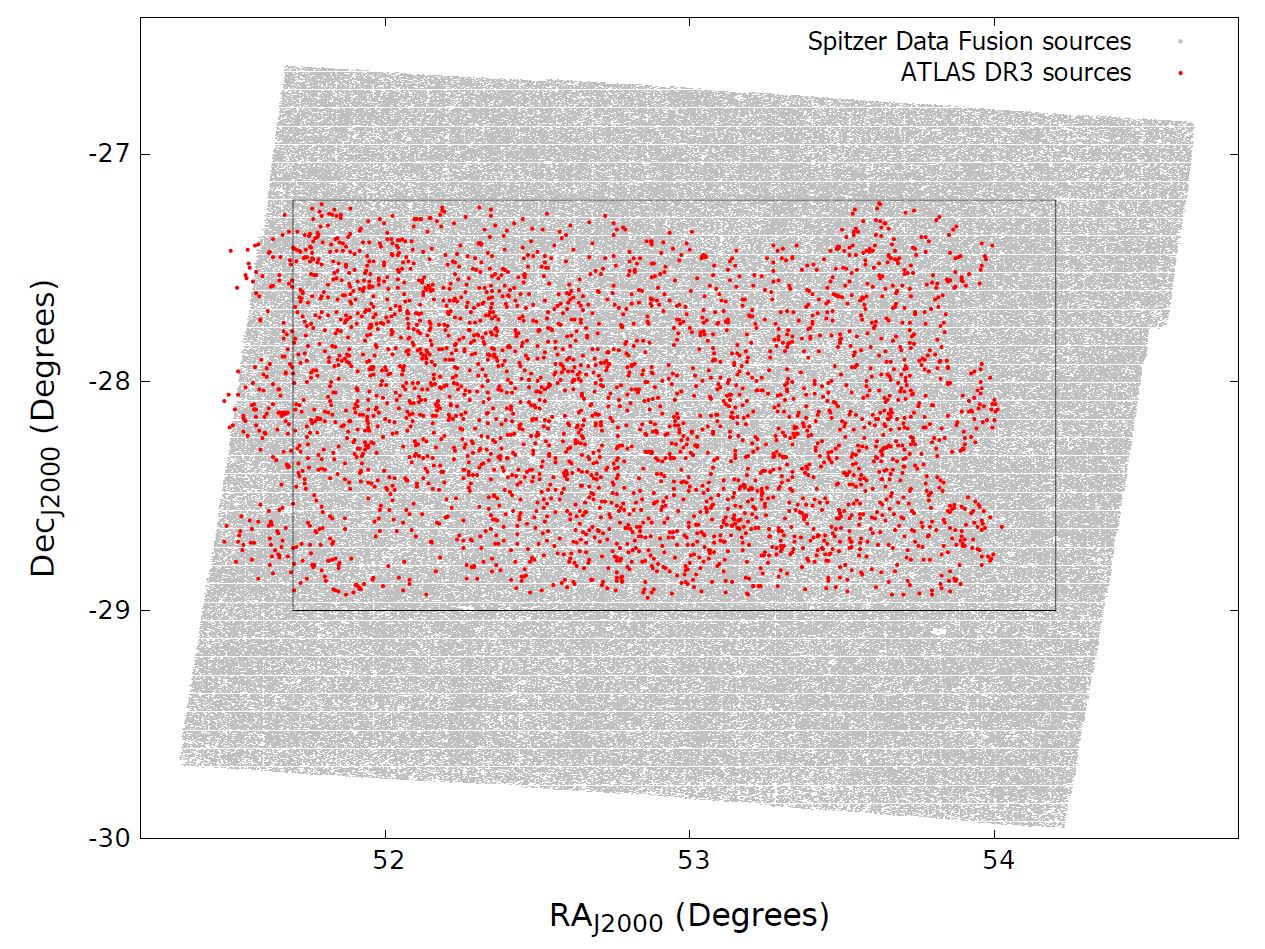}
     \includegraphics[width=0.95\columnwidth]{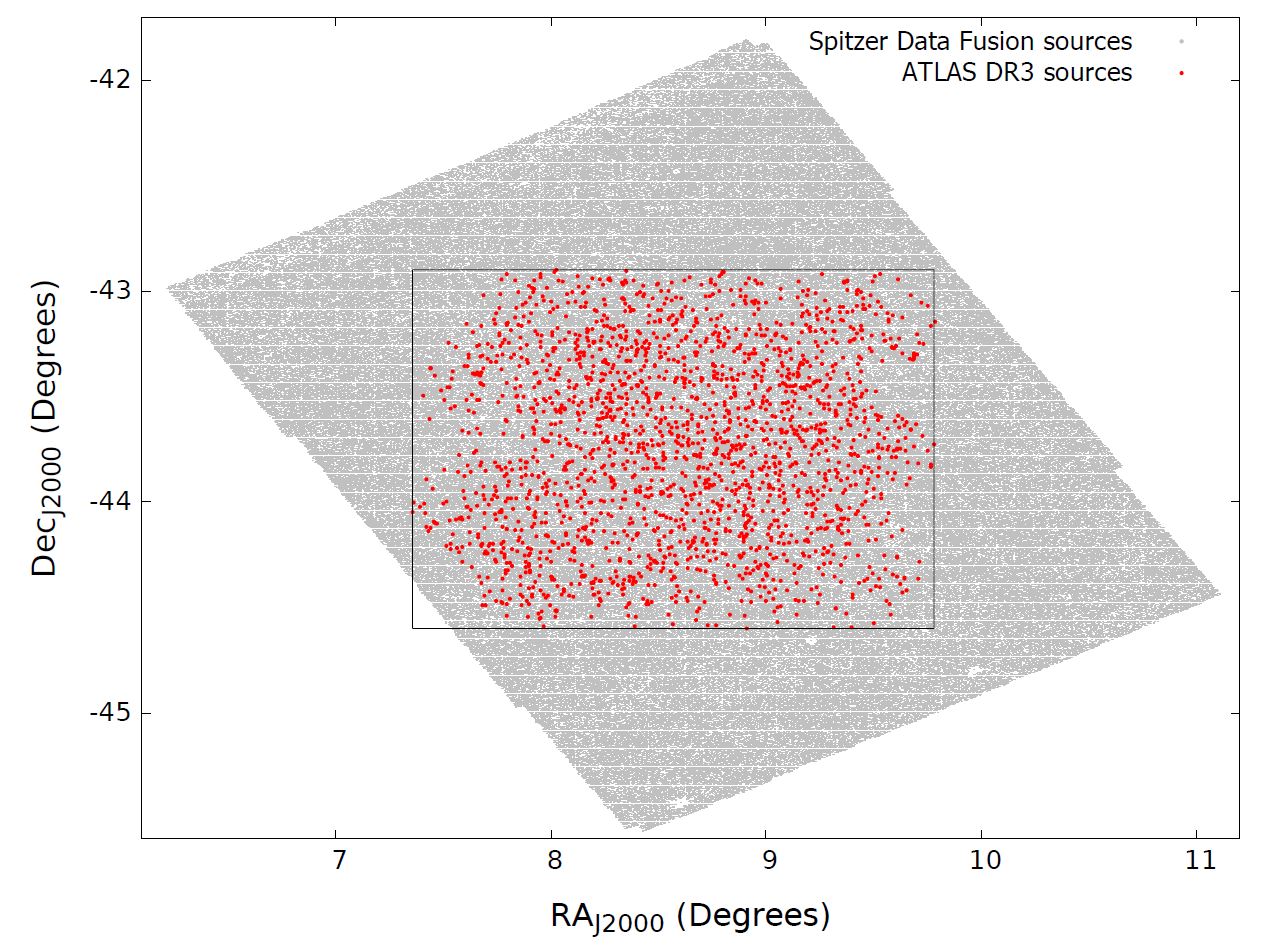}
     \captionof{figure}{Spatial distribution of ATLAS and Fusion sources in CDFS field (top) and in ELAIS-S1 (bottom). We also mark the rectangular regions used in this work. }
     \label{fig:atlas_fusion_overlap}
\end{figure*}

\subsection{Catalogue Coverage}
\label{coverage}
We overlay all the sources in the ATLAS and Fusion catalogues in Figure \ref{fig:atlas_fusion_overlap}. While the Fusion catalogue completely covers the ATLAS observations in ELAIS-S1, part of the ATLAS CDFS data is not covered by the Fusion catalogue. Hence we restrict all our analysis to the following sub-region for CDFS ($51.7 {\degree} \leq RA \leq 54.2 \degree$  and  ${-29.0}  \degree \leq Dec \leq {-27.2} \degree$). This region has been placed so that it is inside Fusion and $100''$ from the edge. We also restrict our analysis in ELAIS-S1 to ($7.3 {\degree} \leq RA \leq 9.7 \degree$  and  ${-44.6}  \degree \leq Dec \leq {-42.9} \degree$).


\section{Source Cross-matching Techniques} \label{scmt}

In this work we further adapt the Likelihood Ratio for the ATLAS and  Fusion catalogues but also extend this technique to account for multiple infrared candidates. In future work we shall show how we adapt it for complex radio sources. 

%

\subsection{Likelihood Ratio Technique}

\citet{Richter:1975} presented an early statistical treatment to cross-match optical sources to the low resolution 5C3 radio survey by applying the statistical separation of real and chance identifications. This technique was then developed by \citet*{Ruiter:77}  to match optical sources to radio sources detected with the Westerbork Synthesis Radio Telescope using a probability ratio, referred to as the Likelihood Ratio (LR). It uses the ratio of the {\it a priori} probability, $dp(r|id)$, that radio source and optical counterpart are intrinsically located at the same position, and the probability that the optical object is an unrelated background or foreground source. This method was further refined by \citet{Sutherland:92} who defined the Likelihood Ratio as the ratio between the probability that a candidate source is the correct identification and the probability that it is an unrelated background or foreground source as a function of magnitude.

The LR technique is commonly used to cross-match low resolution long wavelength surveys with optical data of higher resolution. For example \citet{Ciliegi:03} used this method to find optical counterparts for the VLA 6\,cm Lockman Hole survey. \citet{Ciliegi:05} used the same technique to look for optical and near-infrared (NIR) counterparts for the VLA 1.4\,GHz survey in the VIMOS VLT deep survey. More recently \citet{Smith:11} used the technique with some further refinements to \citet{Sutherland:92} to identify optical counterparts to $250\, \mu$m sources from the {\it Herschel}--ATLAS survey. The refinements from the \citet{Smith:11} technique has been followed by \citet{Fleuren:12} with some modifications when matching sources between the NIR VISTA VIKING and {\it Herschel}--ATLAS SPIRE catalogues. 

In recent applications \citep[e.g. by ][]{Smith:11} this is presented in the form:

\begin{equation} \label{eq:lr}
   L= \frac {q(m) f(r)} {n(m)}
\end{equation}

\noindent where \begin{math} q(m) \end{math} is the probability distribution of the true counterparts as a function of magnitude $m$; \begin{math} f(r) \end{math} is the distribution of probability density per unit solid angle, and \begin{math} n(m) \end{math} is the surface density of background and foreground objects.

In the following sub-sections the terms in Equation~\ref{eq:lr} are discussed with reference to the Fusion catalogue and the ATLAS catalogue. As the Fusion catalogue provides $3.6\,\mu$m flux densities we simply use flux densities, $S_\nu$, rather than magnitudes from this point. 

%

\subsubsection{The Probability Distribution Function}\label{f_r}

Here we follow the standard approach to the definition of the LR \citep{Sutherland:92}. Therefore we use \begin{math} f(r) \end{math} in Equation~\ref{eq:lr} as the probability distribution function (PDF) of the positional errors, see also the definition of \begin{math} f(r) \end{math} given by \citet{Fleuren:12}. We note a confusion in definition of \begin{math} f(r) \end{math} in \citet{Smith:11}, where they first define \begin{math} f(r) \end{math} as ``the radial probability distribution of \textit{offsets} between the 250-μm positions and the SDSS r-band centroid'', that is as the PDF of the offsets between objects of two catalogues, then (in the next paragraph) as the ``probability distribution function of the positional error''. The difference between two definitions is significant, because the ``probability distribution function of the positional errors'' is determined by the Gaussian function, whence the PDF of the offsets between objects of two catalogues is described by the Rayleigh distribution function~\footnote{We thank the reviewer who attracted our attention to this confusion in \citet{Smith:11}}. In our case \begin{math} f(r) \end{math} is a two-dimensional Gaussian distribution of the form: 

\begin{equation}\label{eq:f_r}
  f(r)= {\frac{1}{2\pi \sigma^2}} {\exp\left({\frac{-r^2}{2\sigma^2}}\right)} .
\end{equation}
Here, $r$ is the angular distance (in arcseconds) from the radio source position, and $\sigma$ is the combined positional error
given by:

\begin{equation}
   \sigma = \sqrt{{\sigma_{\rm Posn}}^2 + {\sigma_{\rm Atlas}}^2 + {\sigma_{\rm Fusion}}^2 } .
\end{equation}

\noindent The Fusion absolute position uncertainty, $\sigma_{\rm Fusion}$, is taken as  $0.1''$ \citep{Vaccari2012} and the ATLAS absolute position uncertainty, $\sigma_{\rm Atlas}$, is taken from \citet{Huynh:05}, who argued that the positional accuracy of 1.4\,GHz ATCA observations for $10\sigma$ detections is $0.6''$.

The positional uncertainty term, $\sigma_{\rm Posn}$, of the individual lower resolution ATLAS sources depends on the signal to noise ratio (SNR) and the full-width at half maximum (FWHM) of the radio restoring beam (point spread function in other words). We use the value for $\sigma_{\rm Posn}$ 
as provided in \citet{Ivison:07} and used in \citet{Huynh:05}:

\begin{equation}
   \sigma_{\rm Posn} \simeq 0.6 \left( \frac{FWHM}{ SNR}\right) 
\end{equation}


As the position angle of the restoring beam is small for both fields (see Table~\ref{table:radio_beam}) we can assume it is zero, hence:



\noindent
\begin{figure}
\centering
          \includegraphics[width=0.99\columnwidth]{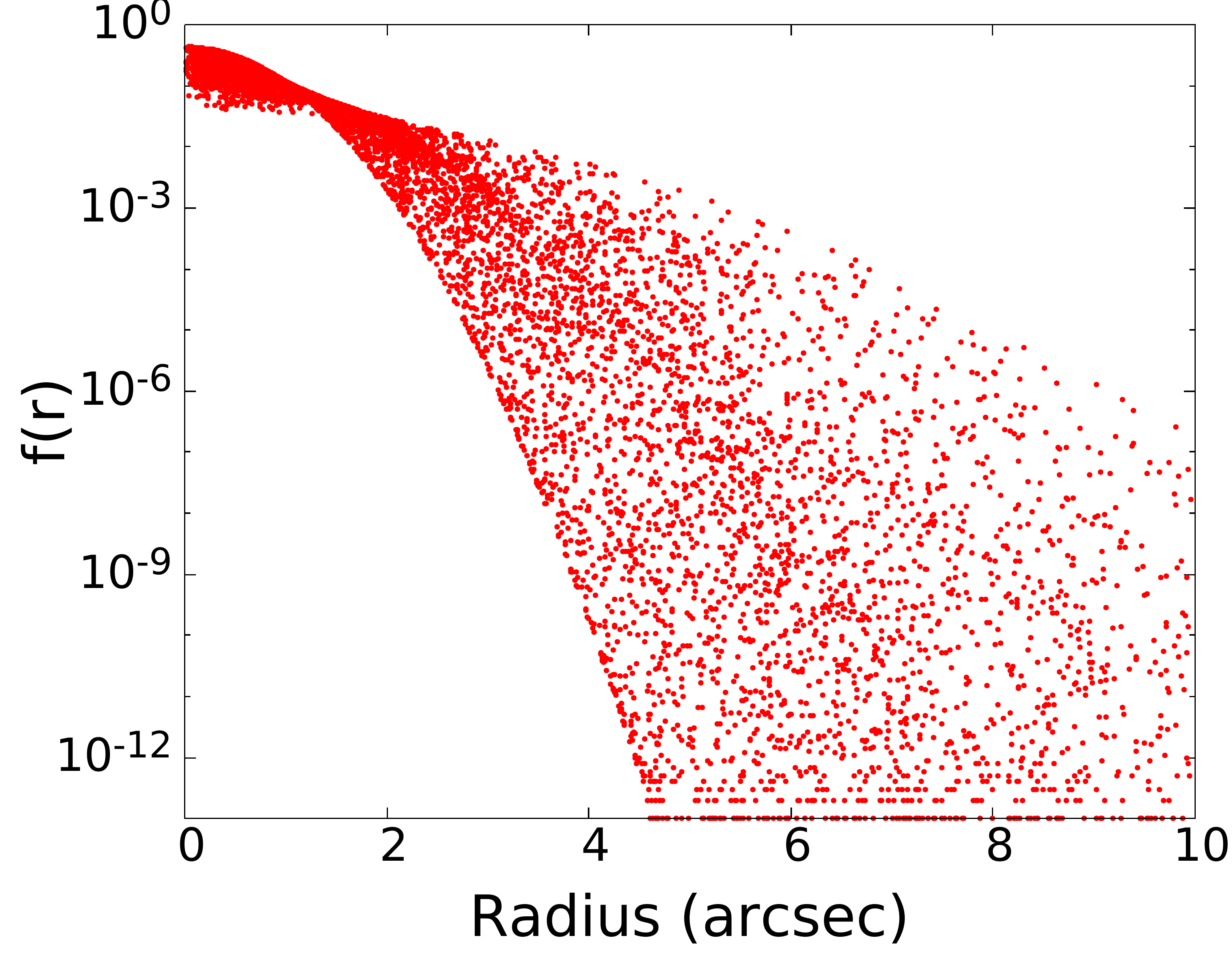}
\captionof{figure}{$f(r)$ vs $r$ for all the potential candidates for both fields CDFS and ELAIS-S1 within the initial $10\,''$ search radius.}
    \label{fig:plot_fr}
\end{figure}

\begin{equation}
   \sigma_{\rm Posn} = \frac{0.6}{SNR} \times \left( \left(\frac{\sin\theta}{\Phi_{\rm Min}}\right)^2+\left(\frac{\cos\theta}{\Phi_{\rm Maj}}\right)^2 \right )^{-1/2}
\end{equation}

\noindent where $\theta$ is the Position Angle of the candidate Fusion counterpart relative to the radio source defined clockwise from North. The SNR values are taken from the ATLAS radio catalogue for each source. The terms $\Phi_{\rm Min}$ and $\Phi_{\rm Maj}$ are the values of minor and major axes of the beam given in Table \ref{table:radio_beam}.


The distribution of the values of $f(r)$ with radius from Equation~\ref{eq:f_r} for the individual candidate Fusion counterparts found within an initial search radius of $10''$, is shown in Figure \ref{fig:plot_fr}. We can see that $f(r)$ is $<10^{-3}$ for $r>6''$. We further discuss the rationale for choosing a final search radius of $6''$ in Section \ref{search_radius}.

%

\begin{figure*}
\centering
         \includegraphics[width=1.0\columnwidth]{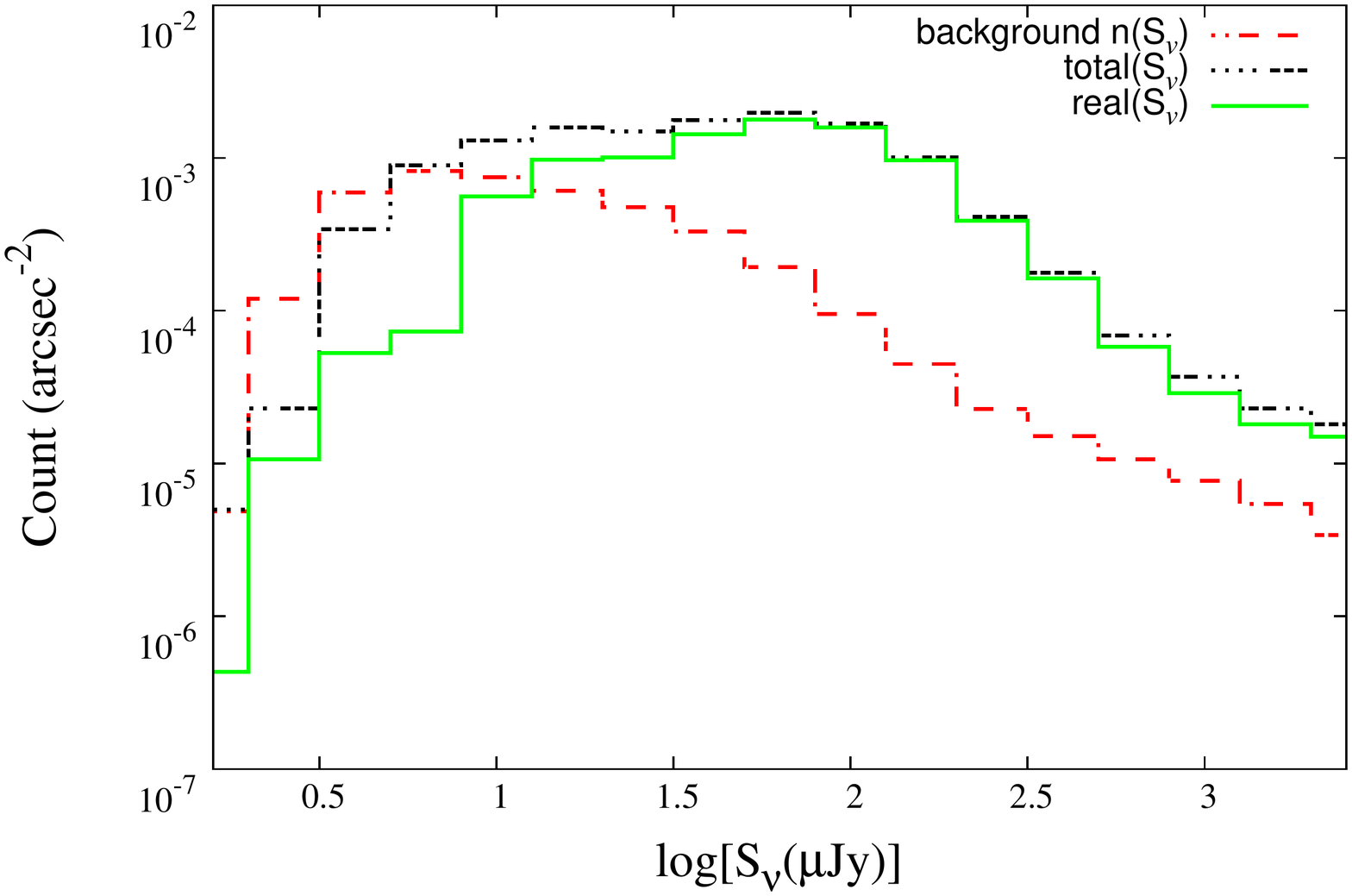}
         \includegraphics[width=1.0\columnwidth]{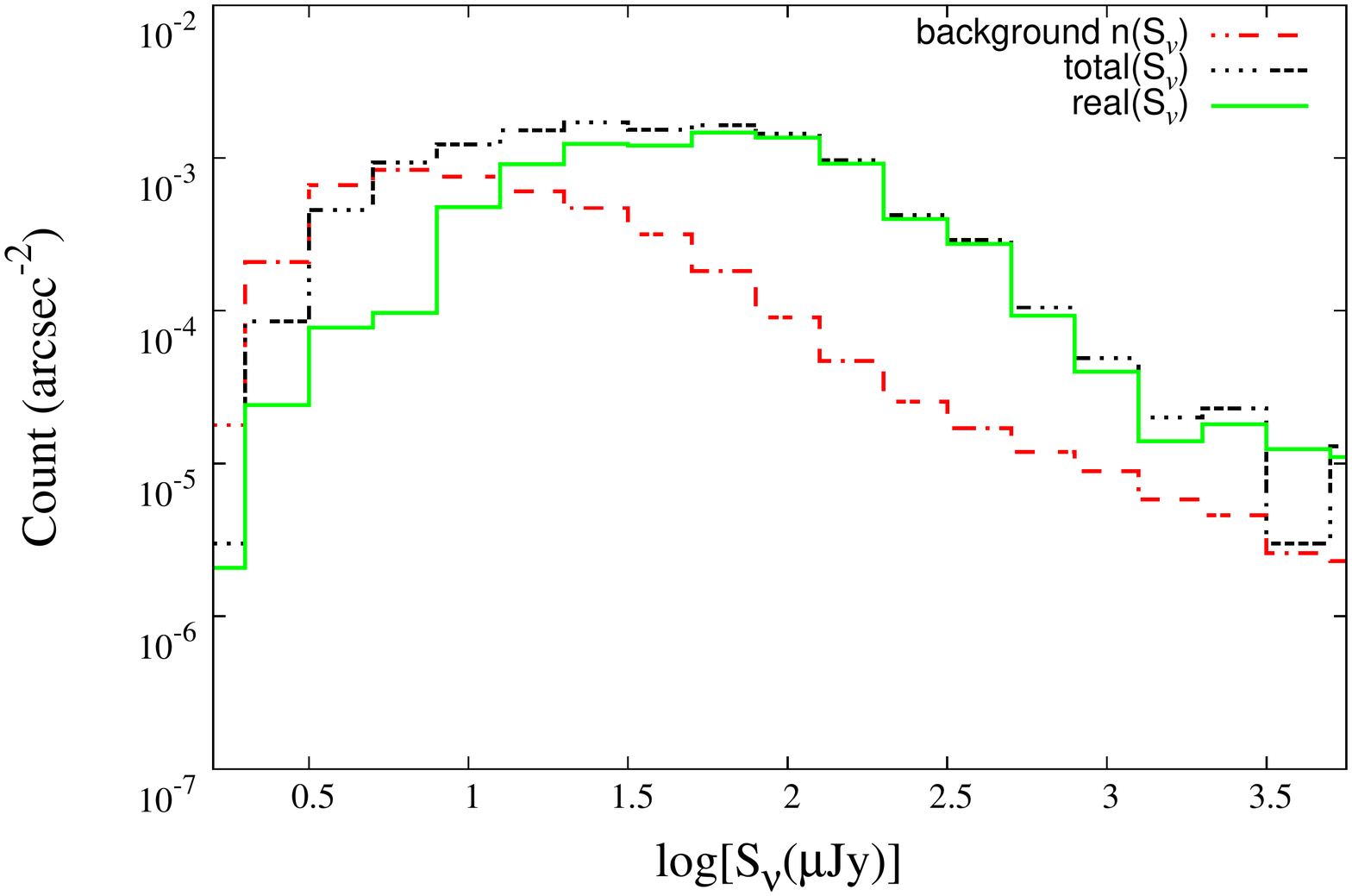}\\
 \captionof{figure}{Histograms of the Fusion values for $n(S_\nu)$ background (red dashed line), $total(S_\nu)$ (black dotted line) and $real(S_\nu)$ (green solid line) for CDFS (top) and ELAIS-S1 (bottom). Note that $S_\nu$ is the $3.6\mu m$ flux.
 }
    \label{fig:both_real_bkgrd_total}
\end{figure*}

\subsubsection{The Background Flux Density Probability Function } \label{background_flux_density}

The quantity $n(S_\nu)$ is the surface density of background and foreground Fusion sources with flux density, $S_\nu$. 
The surface density of Fusion sources not related to ATLAS radio sources can be obtained from the Fusion catalogue by one of two methods, both of which have been implemented within the LRPY algorithm:

\begin{enumerate}[leftmargin=1cm]
     \item Use all Fusion sources within an annulus of $6'' < r < 100''$ around each radio candidate --- this is referred to as the \textit{local} method.
     \item Use all Fusion sources from the area of overlap between the two catalogues (defined in \S~\ref{coverage}) --- this is referred to as the \textit{global} method and is the default in LRPY.
\end{enumerate}

With the \textit{local} method care must be taken that the annuli are not too close to the edge of the field as this can result in a lower count for the background sources as they encompass regions beyond the survey with no sources. To mitigate this edge effect, only annuli $100''$ from the inside edge of the area are used. 
The flux densities are binned and the resultant $n(S_\nu)$ values are then divided by the total area covered to produce a density function, for the CDFS and ELAIS-S1 fields (see Figure \ref{fig:both_real_bkgrd_total}). These values are stored in a database lookup table for use later in the final LR calculations.

There are advantages and disadvantages to both methods: the {\it local} method can account for variations in depth and density of a catalogue, and for very large surveys the entire catalogue is not required, but the {\it global} method can provide better statistics, if area is limited and depth is uniform, which can be important for both bright and faint flux densities where numbers are small. We use the global method as default as it best suits our situation with the Fusion catalogue being uniform in depth.

%
%
\subsubsection{The True Counterpart Probability Distribution  }\label{TTCPD}

The true counterpart probability distribution, $q(S_\nu)$, is the probability that a true Fusion counterpart to a radio source has a flux of $S_\nu$ at $3.6\,\mu m$ :

\begin{equation}\label{eq:q_m}
	q(S_\nu) = {\frac{ real(S_\nu)} {\sum_{S_\nu}real(S_\nu)}} \times Q_0
\end{equation}

\noindent Here $real(S_\nu)$ is the background subtracted distribution of flux densities of Fusion sources around an ATLAS source. The coefficient $Q_0$ represents the probability that a real counterpart is above the detection limit in the matching catalogue; it does not depend on the search radius. To determine $real(S_\nu)$ we take:

\begin{equation}
real(S_\nu)=total(S_\nu)-n(S_\nu) 
\end{equation}

\noindent where $n(S_\nu)$ is the surface density of unrelated background/foreground sources introduced in the previous subsection and  ${total(S_\nu)}$ is the surface density of all Fusion sources to be matched within the search radius, $r$, including the true counterpart (if above the detection limit) plus unrelated background and foreground sources. 
These values are kept in the same LRPY database table as $n(S_\nu)$ for use later by the algorithm. 

The distributions of $real(S_\nu)$ and $total(S_\nu)$, as well as distribution of $n(S_\nu)$ are shown in Figure \ref{fig:both_real_bkgrd_total}. It should be noted that for the unphysical condition where $n(S_\nu) > total(S_\nu)$ (i.e. when the background exceeds the measured distribution), a method is adopted to set $real(S_\nu)$ to be positive. This occurs at faint and bright flux densities when there is a small number of Fusion sources in a given flux density bin. 
To keep our estimate of $real(S_\nu)$ positive and physical, we replace negative values of $real(S_\nu)$/$total(S_\nu)$ with a value determined from the last positive value
at faint and bright flux densities. This adaption ensures we account for potential counterparts at the extreme flux density values. 

\begin{figure*}
\centering
         \includegraphics[width=0.9\columnwidth]{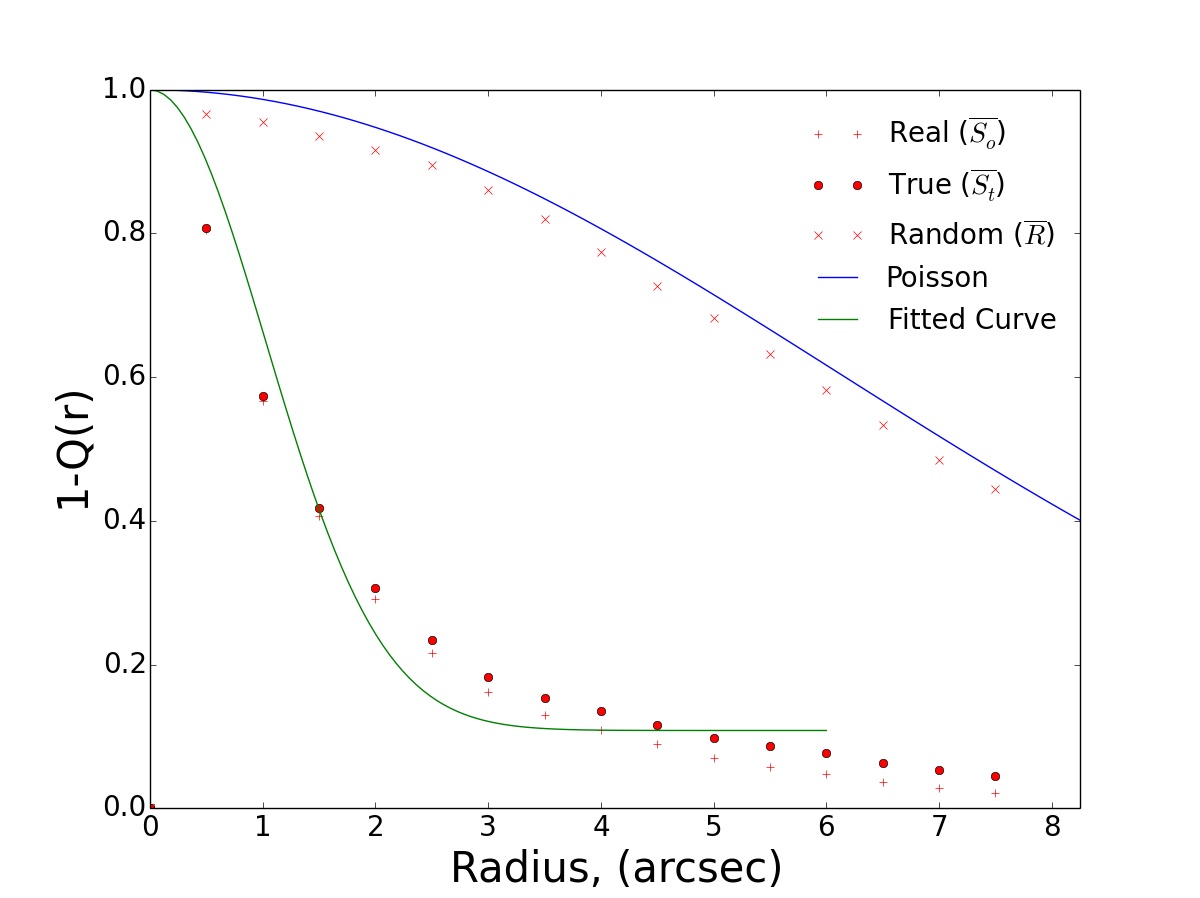}
         \includegraphics[width=0.9\columnwidth]{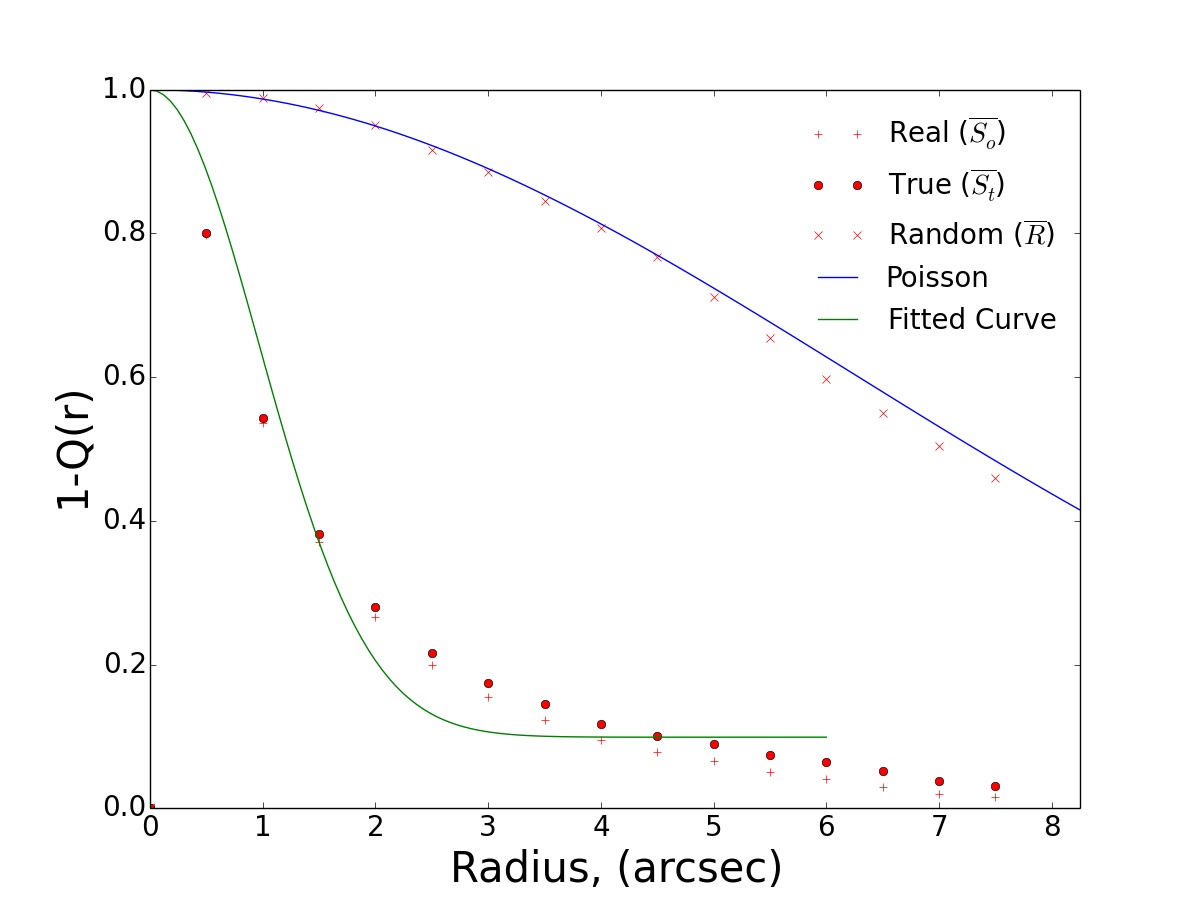}
 \captionof{figure}{Estimation of $Q_0$ for CDFS (top) and ELAIS-S1 (bottom) determined from fitting the ratio, $\overline{S_t}$ (red filled circles), of the fraction of observed blanks, $\overline{S_o}$ (crosses), and the fraction of random blanks, $\overline{R}$ (plusses). The green line represents the functional fit to the ratio (Equation~\ref{eq:q0}), and the blue line is an estimate of the fraction of random blanks from Poisson statistics using Equation~\ref{eq:poisson}. Taking \citep{Fleuren:12}, the dependence of $Q$ on the search radius can be presented in the form $Q(r) = Q_{0}  \exp(-r^2/2\sigma^2)$.}
    \label{fig:both_q0}
\end{figure*}

A reasonably accurate determination of $Q_0$ in Equation \ref{eq:q_m} is naturally required. If we were simply to estimate $Q_0$ by summing $real(S_\nu)$ and dividing by the total number of ATLAS sources we would likely over-estimate $Q_0$ due to source clustering and genuine multiple matches (which we deal with in \S\ref{doubles}). 
While this simple method finds values of $Q_0=0.845$ for CDFS and $0.822$ for ELAIS-S1, we undertake the following process to estimate its value more accurately.
We follow \citet{Fleuren:12} who, to avoid these issues, estimate the value $1-Q_0$, which in this case will be the fraction of ATLAS sources {\it without} a Fusion counterpart, which we refer to here as `blanks'.  These will principally be
ATLAS sources with true counterparts below the Fusion detection limit, or ATLAS sources with true Fusion counterparts outside the search radius. The latter case is possible when ATLAS sources are complex and the Fusion counterpart may well not correspond to any radio component, but lie between components (lobes) which can be separated by tens of arcsec.

The true fraction of blanks, $1-Q_0=\overline{S_t}$, will be greater than the observed fraction of blanks, $\overline{S_o}$, because a fraction of true blanks will have random (i.e. physically unrelated) Fusion sources within the search radius. Hence, we do not wish to falsely count such sources as matches. Therefore, $\overline{S_t}$ equals $\overline{S_o}$ plus some fraction of true blanks `contaminated' by random Fusion sources. Hence, 

\begin{equation}
	\overline{S_t}=\overline{S_o}+\overline{S_t}\times\frac{R}{N}
\label{eq:st}
\end{equation}

\noindent where $R$ is the number of sources, out of $N$ with randomly generated positions, containing one or more Fusion sources within the search radius, and $N$ is the total number of radio sources. If we define $\overline{R}$ as the number of $N$ randomly generated sources which do not have a Fusion counterpart within the search radius, such that $N=R+\overline{R}$, it is straightforward to show:

\begin{equation}
\overline{S_t}=\overline{S_o}+\overline{S_t}\times\bigg(\frac{N-\overline{R}}{N}\bigg)
\end{equation}

\begin{equation}
\frac{\overline{S_t}}{N}=\frac{\overline{S_o}}{\overline{R}}
\label{eq:st2}
\end{equation}

Hence, one can determine the fraction of true blanks, $\overline{S_t}/N$, as a function of search radius, $r$, by determining the ratio of the number of observed blanks, $\overline{S_o}$, to the number of blanks from a randomly generated catalogue, $\overline{R}$, as a function of $r$. We calculate this result for our case by counting the number of observed blanks with increasing search radii across $0''<r<20''$ and repeat for a catalogue of $N$ randomly generated positions of Fusion sources. We present these results in Fig.~\ref{fig:both_q0} showing, as a function of radius, the fraction of observed blanks, $\overline{S_o}/N$, and the fraction of random blanks, $\overline{R}/N$ and their ratio which equals $\overline{S_t}/N$. As the radius, $r$, increases to encompass all true counterparts, this result tends toward $1-Q_0$. We can fit the distribution in Fig.~\ref{fig:both_q0} with the following expression:

\begin{equation}
\frac{\overline{S_t}(r)}{N} = 1-Q_0\times (1-e^{-r^2/2\sigma^2})
\label{eq:q0}
\end{equation}

\noindent from \citet{Fleuren:12} where $\sigma$ is positional uncertainty. This function returns unity at $r=0$ and $1-Q_0$ for large $r$. By fitting for $Q_0$ and taking $\sigma$ as the maximum value for the field ( $\sigma_{CDFS}=1.08$ arcsec, $\sigma_{ELAIS-S1}=0.868$ arcsec ) to the function, using a non-linear least squares fit, we obtain for both these fields the values and uncertainties for $Q_0$ presented in Table~\ref{table:q_0}. These values are fairly similar to the ones from our earlier crude estimate, but with the CDFS being a little higher and ELAIS-S1 being slightly lower. We note that this function must pass through (0,1) by definition, but may deviate within the best match search radius due to physical clustering of sources or from the existence of multiple true components. 


We also note that we can model the distribution of random blanks, $\overline{R}$, in Fig.~\ref{fig:both_q0}. The probability that an observed area of sky, $a=\pi r^2$, has one or more random Fusion source is given by the Poisson distribution $P(a)=1-e^{-a\lambda}$, where $\lambda$ is the density of Fusion sources. Hence, from Equation~\ref{eq:st}, we can write

\begin{equation}
	\overline{S_t}=\overline{S_o}+\overline{S_t}(1-e^{-a\lambda})
\end{equation}

\noindent which we can rearrange to 

\begin{equation}
	\overline{S_t}=\overline{S_o}e^{a\lambda}
\end{equation}

\noindent and therefore from Equation~\ref{eq:st2} we get:

\begin{equation}
	\frac{\overline{R}}{N}=e^{-a\lambda}
    \label{eq:poisson}
\end{equation}

In Fig.~\ref{fig:both_q0} we overlay this function on the random blanks distribution with radius using a density of Fusion sources of $\lambda=0.004$ arcsec$^{-2}$, for both fields. We note this theoretical determination matches our empirical determination well.




\subsubsection{The Search Radius} \label{search_radius}

\citet{Fleuren:12} deal with 1,376,606 near-IR sources in the area of $56\deg^2$, which results in density of near-IR sources of $ \lambda = 6.8\,$ arcmin$^{-2}$ 
and mean intersource distance of $r_0=(\pi \lambda)^{-1/2} \sim 13''$. They chose the search radius $r=10''$, which is 77\% of the mean intersource distance. In our case, the Fusion source density is higher ($\sim15.1\,$arcmin$^{-2}$) and therefore the mean intersource distance is smaller: $r_0=8.7''$. To be consistent with \citet{Fleuren:12} we chose the search radius at 77\% of our mean intersource distance: $r=8.7'' \times 0.77 \sim 6''$. Also as shown in section \ref{f_r} the function $f(r)$ exponentially decreases making the LR vanishingly small, $<10^{-3}$, outside $r=6''$.




\begin{table}
 \centering
  \caption{Estimated fraction, of the non-blanks, $Q_0$ (ATLAS sources with a true counterpart), and its error $\delta Q_0$.}
  \begin{tabular}{@{}lrr@{}}
  \hline
  \hline
   Field & $Q_0$  & $\delta Q_0$ \\
    \hline
    CDFS     & 0.831 & 0.018  \\
    ELAIS-S1 & 0.825 & 0.017 \\
  \hline
  \hline
\end{tabular}
\label{table:q_0}
\end{table}

%

\setlength{\tabcolsep}{4pt} 

\begin{table}
 \centering
  \caption{Statistics of Fusion counterparts inside the $6''$ search radius around ATLAS sources. The first column is the number of Fusion matches; the second column is the number of ATLAS  sources with the corresponding number of Fusions matches (M) for CDFS; the third column is the percentage of the total. Columns four and five are the same, but for the ELAIS-S1 field.}
  \begin{tabular}{@{}ccccccc@{}}
   \hline
   \hline
                          &  & CDFS    &               &                  &  ELAIS  S1     &               \\
    $M$  & $Poisson$ & $Count$ &  $ \% $  & $Poisson$ & $Count$   &  $ \% $    \\
   $(matches)$  &  &  &    &  &    &      \\
   \hline
                  0      &  1905  &   378    & 12.2        &  1307   &   177     &  8.3        \\
                  1      &   914   & 1657    & 53.8        &    628   & 1157     &  54.7        \\
                  2      &   219   &   832    & 27.0        &    151   &   615     &  29.1        \\
                  3      &     35   &   185    & 6.0         &      24   &   138     &   0.6        \\
                  4      &       4   &     23    & 1.0         &        3   &    24      &   1.1         \\
                  5      &       0.4 &     3     & 0.1         &        0.3 &    2       &   0.1       \\
  \hline
                Totals & & 3078      &                &  & 2113       &               \\
  \hline
  \hline
\end{tabular}
\label{table:matches_distribution}
\end{table}

\section{Analysis} \label{Analysis}

In this section we analyse different aspects of the resultant cross-matches and present how we determine criteria for selecting true matches from the LR and reliability values. We then present a method to identify potential pairs of Fusion sources where both are likely contributing to the radio emission of an ATLAS source. 


\subsection{Proposed Selection Criteria} \label{Rel_cutoff}

\begin{figure*}
  \centering
        \includegraphics[width=0.9\columnwidth]{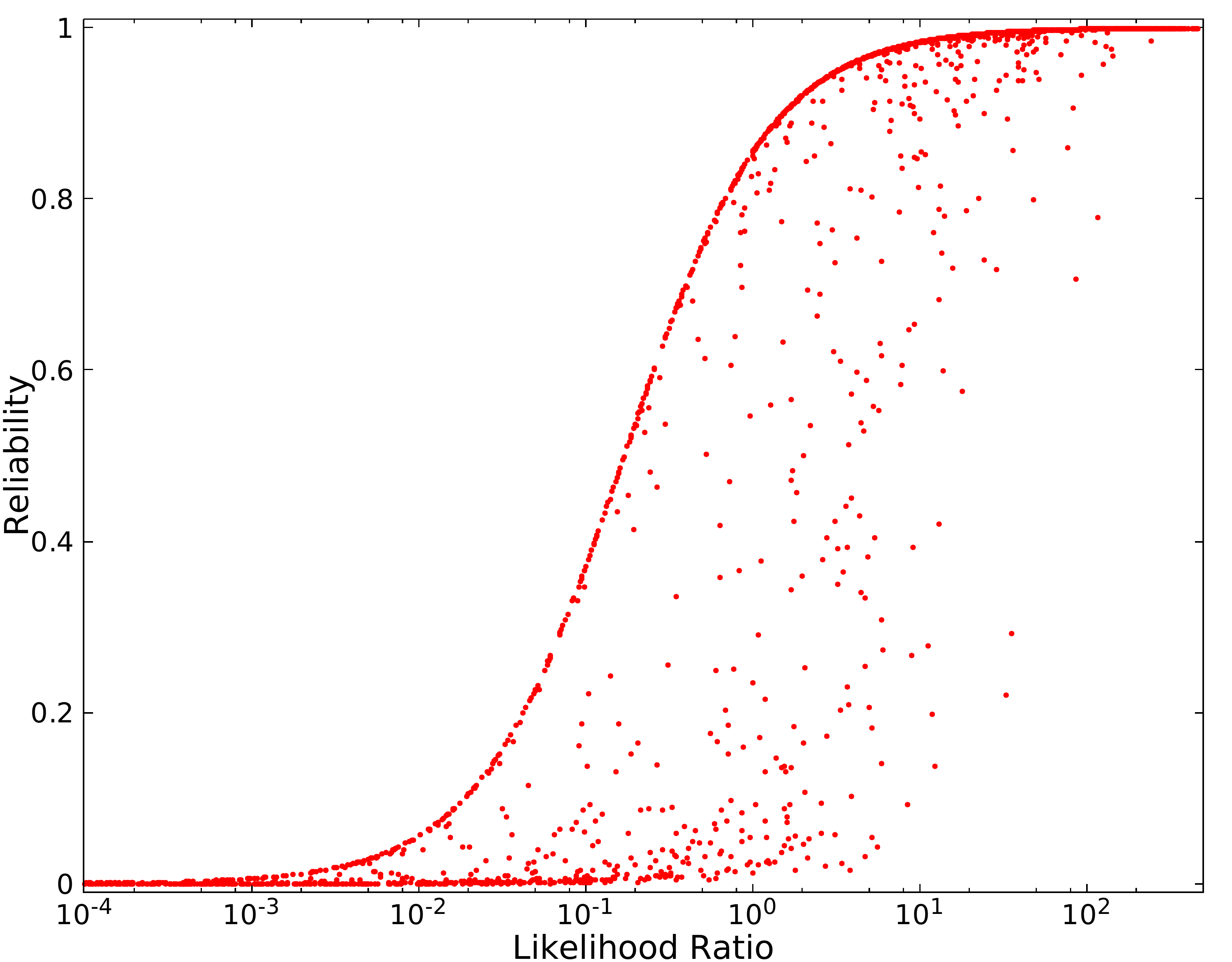}
        \includegraphics[width=0.9\columnwidth]{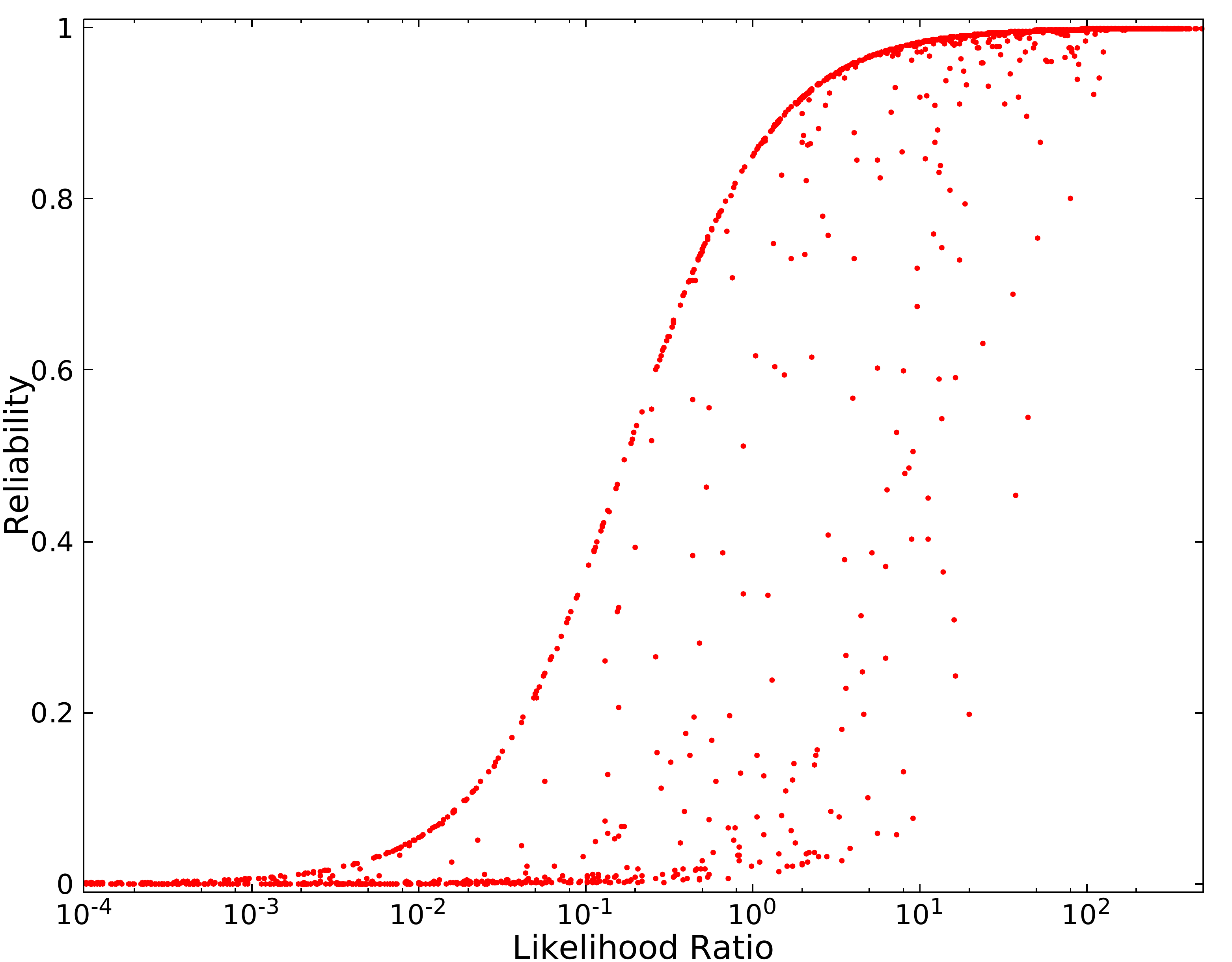}
\captionof{figure}{Plots showing the variation of the reliability, $R$, as a function of the likelihood ratio or CDFS (top) and ELAIS S1 (bottom). For both plots we note some symmetry of points around $R \approx 0.5$ and as discussed in section \ref{ir_doubles} could be used to identify potential Fusion pairs being related to one radio source.}
    \label{fig:rel_vs_lr_plot}
\end{figure*}



Due to the high density of background sources in the Fusion catalogue there can be $0 \leq M \leq 5$ possible candidate Fusion counterparts for a given radio source within the search radius of $6''$ (see Table \ref{table:matches_distribution}). Included in Table 3 is the expected number of radio sources in each field with N fusion potential counterparts from a random distribution, i.e. via Poisson statistics. The numbers we find are higher than those from Poisson statistics suggesting (a) potentially more than one galaxy is contributing to the radio emission and (b) there may be clustering around the host galaxies of radio sources. The former option is discussed in Section \ref{ir_doubles} and we noted the latter point in Section \ref{TTCPD}.



To select from these $M$ possible candidates a reliability value for each can be determined thus:

\begin{equation} \label{equ:reliability}
	R_j = \frac{ L_j} {\sum_{i=1}^{M} L_i + (1-Q_0)}
\end{equation}

\noindent where $R_j$ is the reliability that the candidate Fusion counterpart $j$ of $M$ possible counterparts is associated to the radio source. The sum is taken over all $M$ potential candidates within the $6''$ search radius and $Q_0$ is the probability that the true Fusion counterpart is above the detection limit (determined in \S\ref{background_flux_density} and presented in Table~\ref{table:q_0}). Plots of Reliability versus Likelihood Ratio for each candidate counterpart for both fields are presented in Figure \ref{fig:rel_vs_lr_plot}.

There is always a trade-off between maximising the number of radio sources with `reliable' counterparts and minimising the contamination of false associations. Equation ~\ref{equ:reliability} permits us to compare the relative likelihood of an association between an ATLAS and a Fusion source in the situation where we have two or more potential counterparts. Determining the appropriate cut-off values in LR and Reliability is therefore crucial for any scientific analysis. 
  
Reliability can also be calculated for the case of a single Fusion source, $M=1$ (one Fusion source in the search radius): 

\begin{equation} \label{equ:reliability_single}
	R_j = \frac{ L} {L + (1-Q_0)} .
\end{equation}

Hence, once a LR cut-off, $L_c$, is determined, the corresponding cut-off value of reliability, $R_c$, can easily be calculated for single sources as we know $Q_0$ (here we take $Q_0=0.85$).  

\begin{figure*}
         \includegraphics[width=0.5\columnwidth]{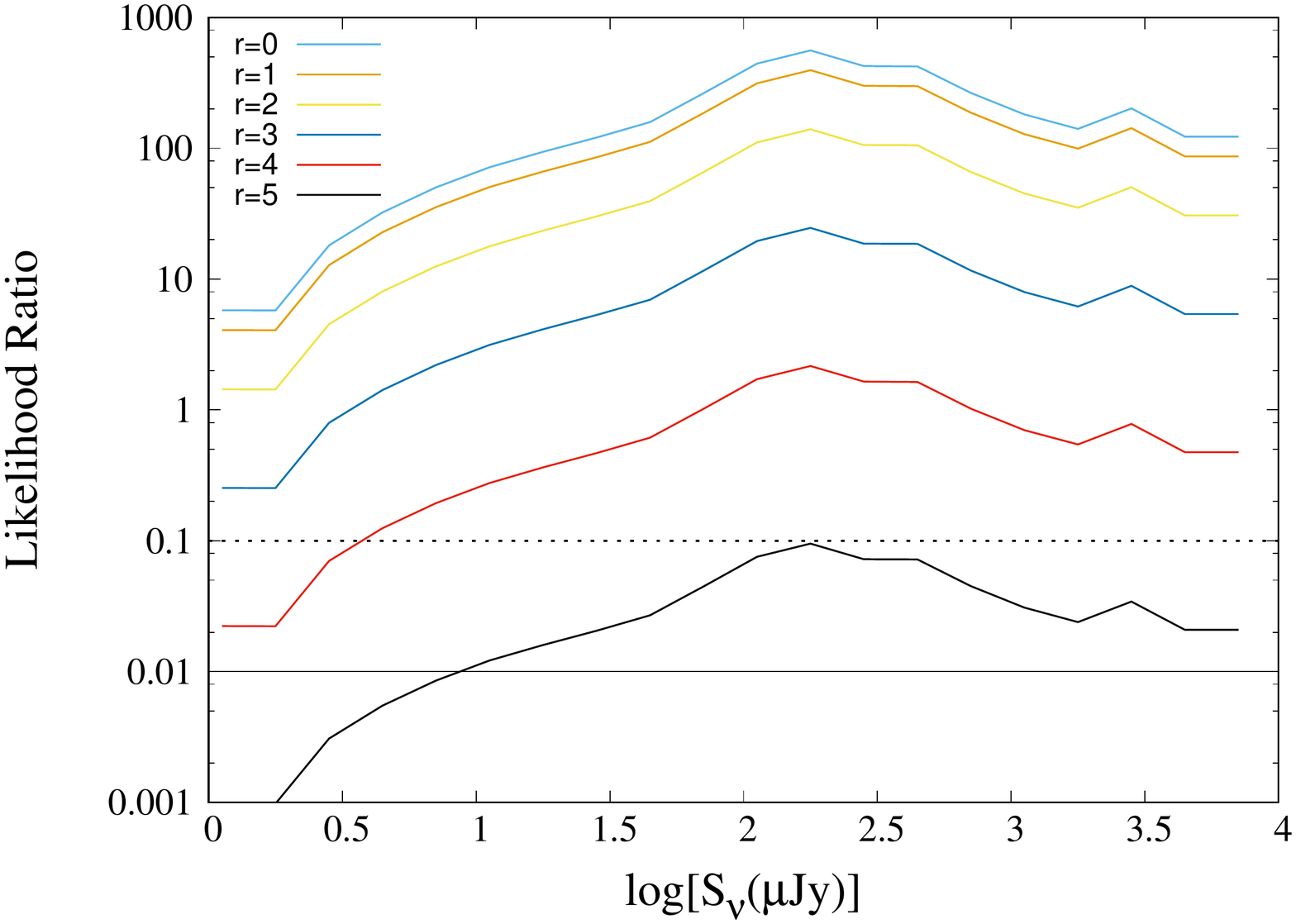}
         \includegraphics[width=0.5\columnwidth]{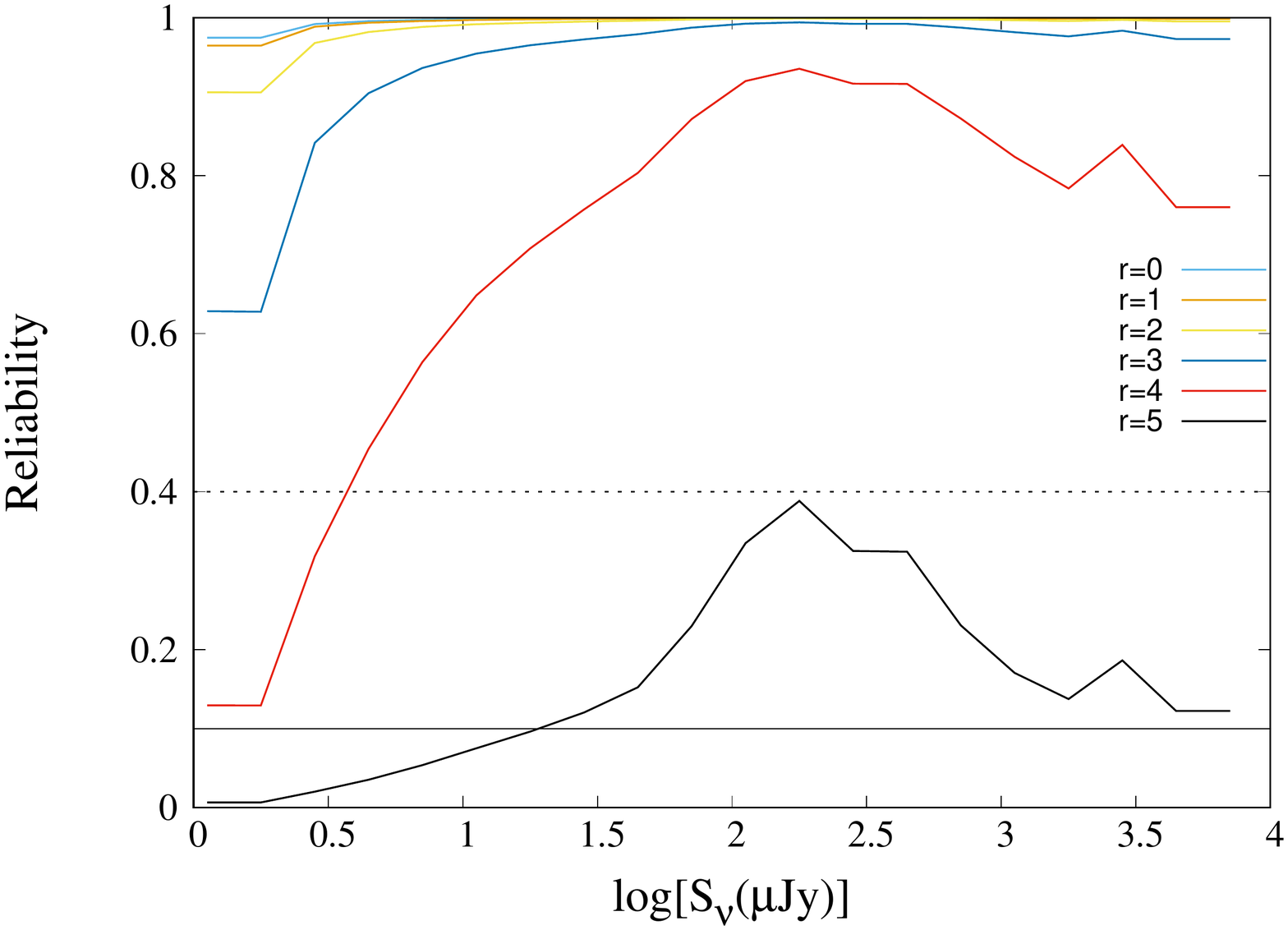}\\
         \includegraphics[width=0.5\columnwidth]{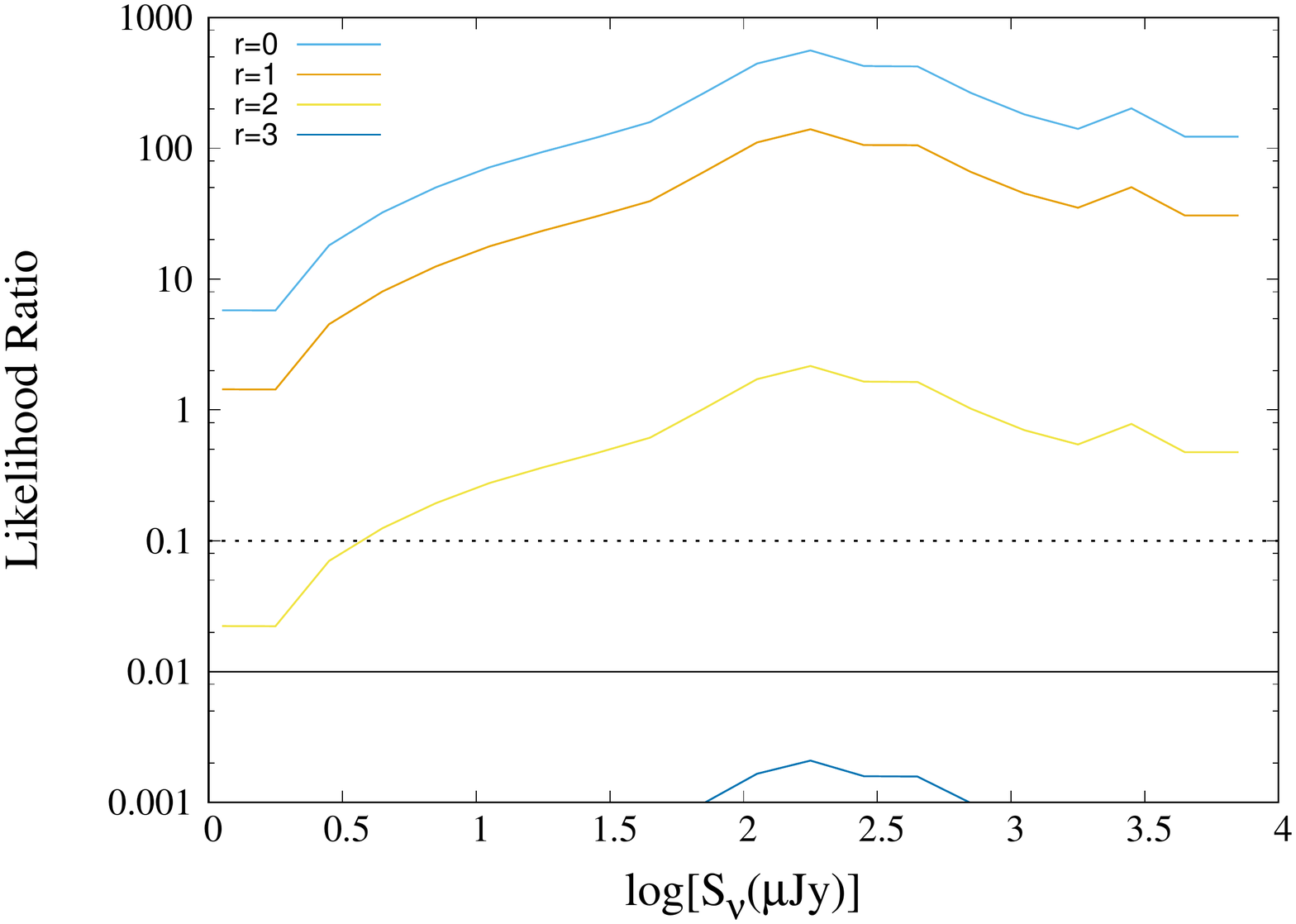}
         \includegraphics[width=0.5\columnwidth]{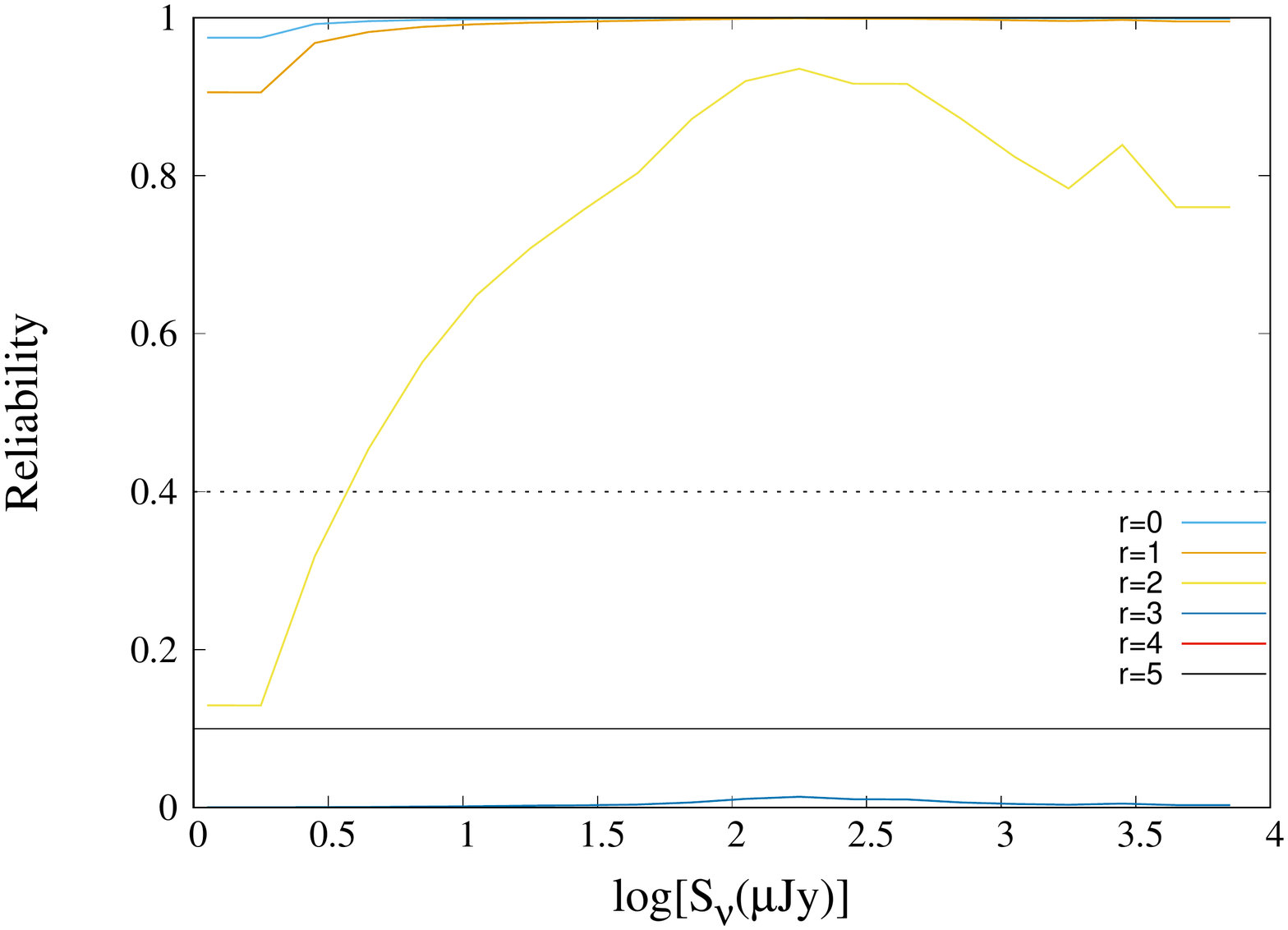}
    \captionof{figure}{Likelihood Ratio against $\log(S_\nu)$ (left column) and Reliability  against $\log(S_\nu)$ (right column) for extreme values of the positional uncertainty $\sigma = 0.6''$ (bottom row) and $1.2''$ (top row) in CDFS. Families of curves are computed for distances $r$ between a candidate Fusion counterpart and the ATLAS source in the range $r=0''$ to $5''$. Distributions of $real(S_\nu)$ and $n(S_\nu)$ used to determine the likelihood ratio are taken from the CDFS field statistics. Horizontal solid lines corresponding to suggested cut-off values presented in \S~\ref{Rel_cutoff}, for $L=0.01$ and $R=0.1$ are drawn on the figures. The horizontal dashed lines show a much stronger selection  criteria $L=0.1$ and $R=0.4$. Note that $S_\nu$ is the $3.6\mu m$ flux.}
    \label{fig:LR_REL_Radius_M_sigma}
\end{figure*}

Figure \ref{fig:LR_REL_Radius_M_sigma} shows the families of theoretical curves $L$ vs. $S_\nu$ and $R$ vs. $S_\nu$ for the range of $r$ (distance between the radio source and Fusion candidate) from $0''$ to $5''$ (all inside the search radius of $6''$) and $Q_0 = 0.85$.  They are calculated for the set of $real(S_\nu)$ and $n(S_\nu)$ we observe in the CDFS field for radio sources detected. The upper plots are computed for $\sigma = 1.2''$, which is close to maximum value of $\sigma$ we deal with in CDFS field (\S 3.1.1), the bottom plots correspond to $\sigma = 0.6''$ (close to minimum value of $\sigma$ in CDFS field).   

We can choose the $L_c$ for single Fusion sources in such a way that for $\sigma=1.2''$ almost all single Fusion sources within $r=5''$ are considered as true counterparts. This condition is fulfilled when $L_c=0.01$ (horizontal dashed line in Fig.~\ref{fig:LR_REL_Radius_M_sigma}) and corresponds to a reliability cutoff of $R_c=0.055$ for CDFS and $0.054$ for ELAIS-S1. These cut-off values are shown in graphs with horizontal solid lines. The horizontal dashed lines show a much stronger criterion for cut-off values of $L_c=0.1$ and the corresponding $R_c=0.37$ for CDFS and $0.36$ for ELAIS-S1. In this case all Fusion sources with $r>4''$ are excluded from consideration as possible counterparts.

Another approach to determining a value for the reliability cut-off, where those candidates with a reliability greater than $R_c$ can be treated as true counterparts, was used by \citet{Smith:11} who estimated the number of false cross-matches using :

\begin{equation} \label{eq:n_false}
     N_{\rm false}(R_c) = \sum_{R^{Max}_{i} \geqslant R_{c}} ( 1-R_{i} )
\end{equation}

\begin{figure}
    \includegraphics[width=0.9\columnwidth]{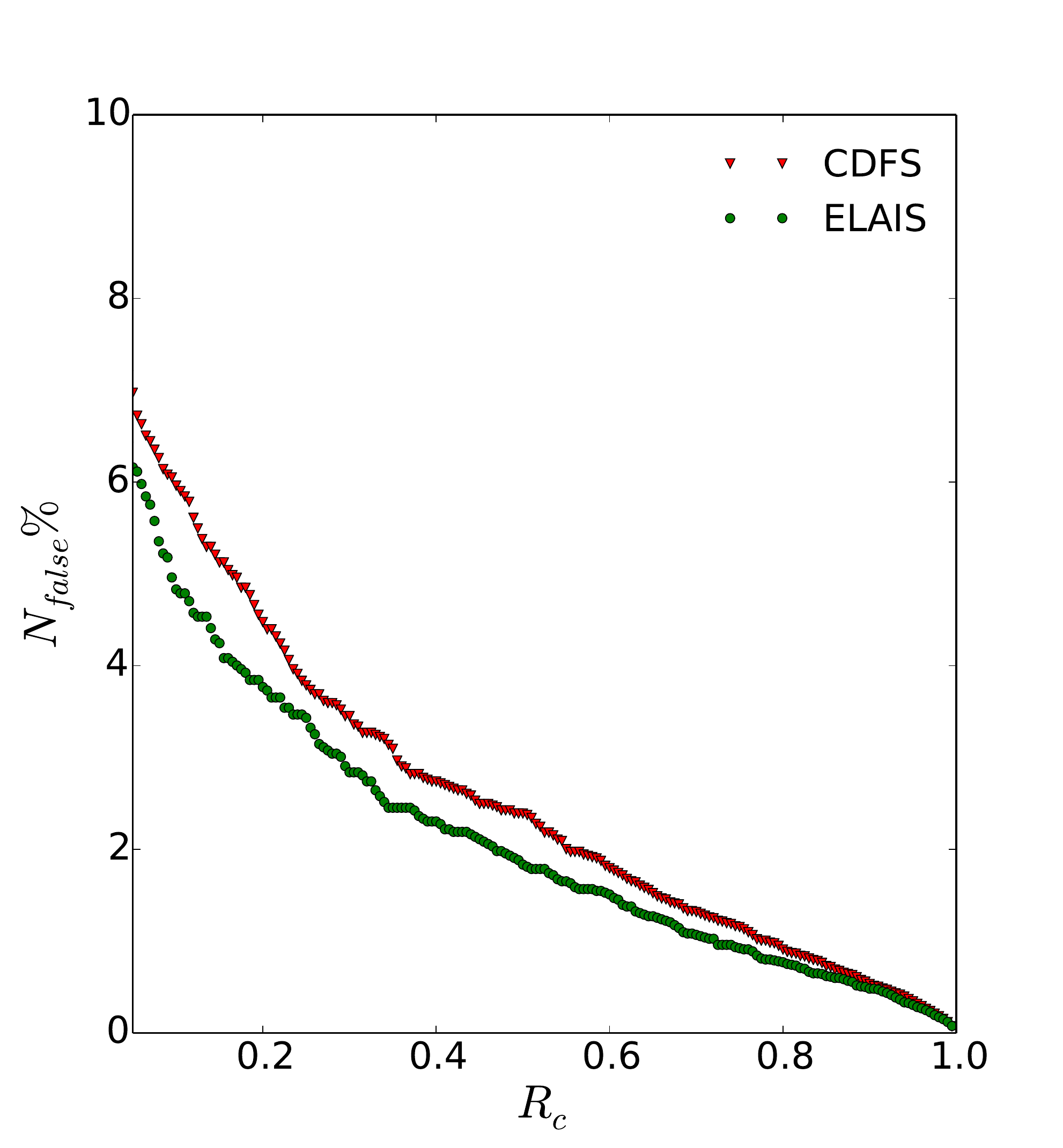}
    \captionof{figure}{Estimated percentage of the false cross-matches, $N_{false}$, as a function of the Reliability cut-off, $R_c$, for CDFS (red) and ELAIS-S1 (green) determined with Equation~\ref{eq:n_false}.}
    \label{fig:both_n_false}
\end{figure}

Figure \ref{fig:both_n_false} shows $N_{\rm false}$ as a function of $R_c$ for our two fields. \citet{Smith:11} used a Reliability limit of 0.8 which gave them a contamination rate of $4.2\,\%$. \citet{Bonzini:12} selected only those candidates with a reliability greater than 0.6 as the threshold to ensure the expected number of spurious associations was below $5\,\%$ of the auxiliary catalogue; and at the same time maximising the number of identified sources. Using a similar acceptable contamination threshold at $5\,\%$ for our datasets, results in $R_c=0.1$ for both CDFS and ELAIS-S1 fields (Figure \ref{fig:both_n_false}). As we discussed above, this value of $R_c$ corresponds to $L_c=0.01$ for single Fusion sources.

Using the $5\,\%$ contamination threshold we can accept only Fusion counterparts
with $L\geq L_{c}$. Here we use $L_{c}=0.01$ and we reject all Fusion counterparts below this value.
We apply this to all Fusion sources, whether
they are single or multiple.

For the ATLAS fields using a LR cutoff of 0.01 (reliability cutoff of 0.1) and using Equation~\ref{eq:n_false}, we have for CDFS $N_{false} = 159$ which is $5.2\,\%$; and for ELAIS-S1 we have $N_{false} = 99$ which is $4.8\,\%$. Using this cutoff there are 2135 ATLAS sources with at least one match in the CDFS field  and 1580 in the ELAIS-S1 field. We give an example of this in the figure~\ref{fig:AppendixIRD_CI0418}.



\subsection{Double and Multiple Fusion Counterparts}\label{ir_doubles}
\label{doubles}

One ATLAS radio source due to its unresolved peak in a low-resolution radio image could potentially be produced by two or more radio sources blended into one apparent ``source'' by the large radio beam. In this section we modify the LRPY algorithm to identify possible double blended radio sources using the background sources from Fusion.

When reviewing the Reliability vs Likelihood Ratio plots in Figure  \ref{fig:rel_vs_lr_plot} we note a symmetry of some data points at high values of the Likelihood around $R\approx 0.5$. This symmetry has also been noted by \citet{Smith:11} in a similar area in their analysis and they surmise that these could be interacting galaxy counterparts; four of their sources had multiple counterparts with spectroscopic redshift differences of $ \Delta z \lesssim 0.001 $. Also \citet{Fleuren:12} highlighte these possible multiple counterparts, and propose that these could be either merging galaxies or members of the same cluster. \citet{Fleuren:12} found matches to $37$ sources (out of 1444) with a mean redshift difference of $0.0011$ with a maximum difference of $\Delta z=0.0187$. Within our catalogue we found these ATLAS sources with two candidate Fusion counterparts to have similar flux density and similar angular separation. Thus we consider the possibility that these pairs of Fusion counterparts could be close or interacting galaxies and may both be contributing to the radio emission from one source. We investigate this further below and group them together introducing the term InfraRed Double (IRD).

When two Fusion sources with similar LRs are found in the search field around a radio source, the reliability of both sources is determined by

\begin{equation} \label{eq:reliability_symmetry}
R=\dfrac{L}{(1-Q_{0})+L/0.5}\;,
\end{equation}

\noindent which follows from Equation~\ref{equ:reliability_single} when $L=L_1=L_2$.  Equation~\ref{eq:reliability_symmetry} results in $R=0.5$ when $L\gg1-Q_{0}$.

\begin{figure}
         \includegraphics[width=0.99\columnwidth]{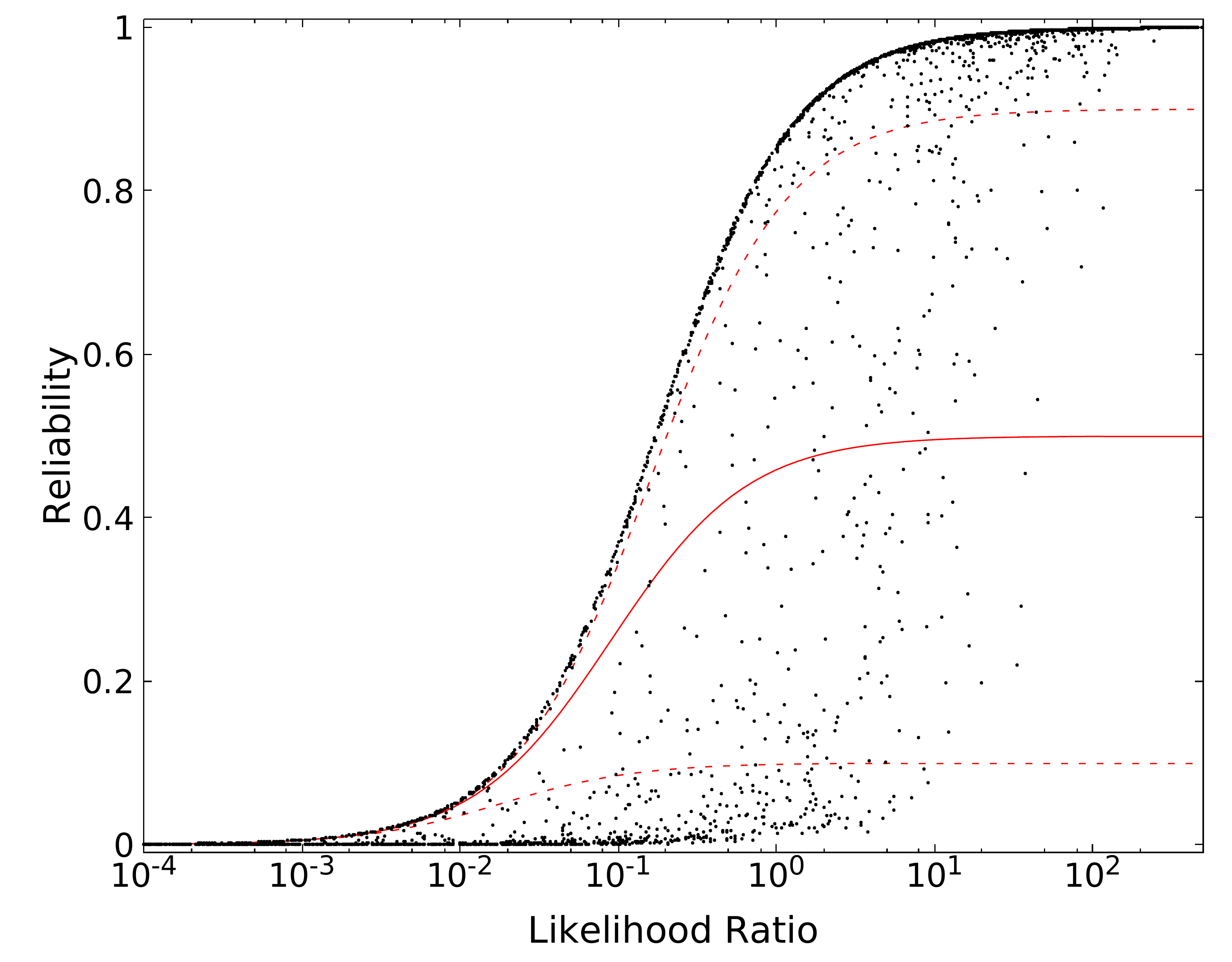}
    \captionof{figure}{Likelihood Ratio vs Reliability for all possible matches within the $6\,''$ search radius for both fields. Also included are the selection limits using Equation~\ref{eq:rel_sym}, the upper and lower selection limits ($\beta \pm 0.4$) are marked with red dotted lines and the axis of symmetry of points ($\beta = 0$) is marked by a solid red line.}
    \label{fig:symmetry_rel_vs_lr}
\end{figure}

In Figure \ref{fig:symmetry_rel_vs_lr} the axis of symmetry for pairs is shown with the red solid curve which follows Equation \ref{eq:reliability_symmetry}. The red dashed curves above and below the axis of symmetry are given by:

\begin{equation} \label{eq:rel_sym}
R=\dfrac{L}{(1-Q_{0})+L/(0.5\pm\beta)}\;.
\end{equation}


\noindent where in this case $\beta=0.4$, so that when $L\gg1-Q_{0}$ these tend toward $R=0.1$ and $0.9$ for the dashed lines. We may then make the hypothesis that if both counterparts have $0.1\leq R\leq 0.9$ then they might both be counterparts Fusion sources and both be contributing to the radio emission. However, if one counterpart has $R<0.1$ then we consider the other counterpart to be the sole true match.


For example, if we have two sources inside the search radius, one with $R_{1}=0.05$ and the other with $R_{2}=0.95$ (so $R_{1}/R_{2}<1/19$), we reject the first source, even if $L_{1}>L_{c}$, and consider the second Fusion source as a single source and sole counterpart. Hence all components of pairs below the lower dashed line are rejected, and all Fusion sources above the upper dashed line are now considered as singles. 
For this work we take a value of $\beta=0.4$ based on the LR and Reliability cut-off values in Figure \ref{fig:LR_REL_Radius_M_sigma}. This acceptance zone can be narrowed or widened by decreasing or increasing $\beta$ in the algorithm.

We have a relatively small subset of multiple Fusion counterparts between the dashed lines in Figure \ref{fig:rel_vs_lr_plot} and with $L>0.01$, but clearly if this selection is applied to much larger catalogues then a significant number of sources would be selected as such. In our case, we have $38$ pairs of fusion counterparts in the CDFS field and $26$ in ELAIS-S1 which makes $\approx2\%$ of all radio sources with cross-identification in Table \ref{table:results}. Hence, from the $\num{7e7}$ radio sources expected in EMU we might expect more than a million with multiple matches.

Using the selection rules as outlined earlier in this section, we find 64 pairs for the two fields. To explore the possibility that some of these 64 pairs of galaxies could be members of the same group of galaxies or even physically interacting, we perform a nearest-neighbour match of the Fusion sources with objects from the OzDES survey (presented in Section \ref{ozdes})


 If the Fusion source is within $1''$ of an OzDES object, we consider it to be the same object. We found 22 out of 64 doubles to have spectroscopic redshifts of {\em both} galaxies which we present in Tables \ref{table:ir_cdfs_doubles_fusion_z} and \ref{table:ir_elais_doubles_fusion_z}.  In 20 cases both objects have similar redshifts, $\Delta z/z < 0.01$, and in two cases Fusion sources have significantly different redshift. 
Postage stamp images of some IRDs are given in Figure \ref{fig:AppendixIRDoublesPS} in the Appendix with the ATLAS radio contours overlaid on the greyscale IR images to demonstrate these objects.

In addition, using the Great Observatories Origins Deep Survey \citep[GOODS][]{Dickinson:2003, Renzini:2003}, we find two of our CDFS IRDs have Hubble Space Telescope (HST) archive images 
\footnote{Based on observations made with the NASA/ESA Hubble Space Telescope, obtained from the data archive at the Space Telescope Science Institute. STScI is operated by the Association of Universities for Research in Astronomy, Inc. under NASA contract NAS 5-26555.}. 
In the Appendix are Figures \ref{fig:AppendixIRD_CI1036} and \ref{fig:AppendixIRD_CI0418}
in which we present these images with the IR source positions marked and the radio contours overlaid. As well as the ATLAS radio contours we present contours from the deep JVLA 1.4\,GHz survey of this sub-region of the CDFS \citep{Miller:13}. The {\it HST} images clearly indicated that these two pairs of galaxies are interacting via their disturbed morphologies and tidal tails.

It is possible that more than two Fusion sources high enough LR and reliabilities above the cut-off. In this work we do not have a situation of this sort. However, when dealing with large data sets, we can expect cases of multiple counterparts, and an automated approach has to be elaborated for this case. 

\begin{center}
\begin{table}
 \centering
  \caption{Results of cross-identification of ATLAS sources with FUSION sources using the LRPY code. We present the total number of radio sources, the number having Fusion coverage, the number with any Fusion counterpart within $6''$ and the number with high reliability Fusion counteparts per Section \ref{Rel_cutoff} and \ref{ir_doubles}.}
  \begin{tabular}{@{}cccc@{}}
   \hline
   \hline
 Field & CDFS & ELAIS-S1 & both\\
 \hline
 Total & 3078 & 2113 & 5191\\
 with Fusion coverage & 2922 & 2113 & 5035 \\
 with any Fusion candidates & 2700 & 1936 & 4636\\
 with high reliability XID & 2222 & 1626 & 3848\\
  \hline
  \hline
\end{tabular}
\label{table:results}
\end{table}
\end{center}

\begin{center}
\tabcolsep=1pt
\begin{table}
 \centering
  \caption{Redshifts for possible IR doubles taken from OzDES, by a nearest-neighbour match between Fusion {\it Spitzer} and OzDES \citep{Yuan:2015} within $1''$. Of these 22, 20 have pairs of galaxies with $\Delta z/z_{\rm spec}<0.01$.}
{\renewcommand{\arraystretch}{1.1}%
  \begin{tabular}{@{}cccccc@{}}
   \hline
   \hline
   ATLAS  &   Fusion    & ang sep    & OzDES                               &  OzDES     &  OzDES                 \\
   ID     &   ID       & (arcsec)         &  ID                                    & z             & $\Delta z$                \\
   \hline
   \multirow{2}{*}{CI0069}   & 309081    &  0.528   & 281939.8                        & 0.6789   & \multirow{2}{*}{0.0019}    \\
                             & 309075    &  0.045  & 00036776      
         & 0.68084  &                                           \\[1.5ex]
   \multirow{2}{*}{CI0099C2}   & 295215    &  0.106   & 0076-01223                        & 0.3339   & \multirow{2}{*}{0.0004}    \\
                             & 295098    &  0.285  & 2940685175      
         & 0.3344  &                                           \\[1.5ex]
   \multirow{2}{*}{CI0175}   & 333146    &  0.448   &91-274837.9                        & 0.1816   & \multirow{2}{*}{0.0001}    \\
                             & 333165    &  0.223  & 32564      
         & 0.1817 &                                           \\[1.5ex]
   \multirow{2}{*}{CI0191}   & 467746    &  0.321   &S117                                & 0.0909   & \multirow{2}{*}{0.0001}    \\
                             & 467716    &  0.121  & NAO\_0552\_119829      
         & 0.09078 &                                           \\[1.5ex]
   \multirow{2}{*}{CI0548}   & 322386    &  0.188  & 57-280213.3
         & 0.5368   & \multirow{2}{*}{0.0004}    \\
                             & 322361    &  0.236  & 2940877666      
         & 0.53631 &                                           \\[1.5ex]
   \multirow{2}{*}{CI0561}   & 151844    &  0.013   & 20-283323.1
         & 0.3402   & \multirow{2}{*}{0.006}    \\
                             & 151810    & 0.106   & 0082-01440      
         & 0.3462 &                                           \\[1.5ex]
   \multirow{2}{*}{CI0632}   & 328609    &  0.231   & NOAO\_0334\_R126091
         & 0.32776   & \multirow{2}{*}{0.0006}    \\
                             & 328657    &  0.076   & NOAO\_0552\_126052     
         & 0.32708 &                                           \\[1.5ex]
   \multirow{2}{*}{CI0633}   & 197618    &  0.277   & S477
         & 0.2511   & \multirow{2}{*}{0.0005}    \\
                             & 197687    &  0.258   & 63053     
         & 0.2516 &                                           \\[1.5ex]
   \multirow{2}{*}{CI0757}   & 171555    &  0.1063   & 0084-00302
         & 0.6633   & \multirow{2}{*}{0.0067}    \\
                             & 171511    &  0.1062   & 0085-00883
         & 0.6701 &                                           \\[1.5ex]
   \multirow{2}{*}{CI1000}   & 178274    &  0.149   & 2939983811
         & 0.3380   & \multirow{2}{*}{0.0002}    \\
                             & 178269    &  1.070   & 49-275932.3
         & 0.3378 &                                           \\[1.5ex]
   \multirow{2}{*}{CI1036}   & 183594    &  0.072   & 92922
         & 1.0967   & \multirow{2}{*}{0.0007}    \\
                             & 183614    &  0.119   & 49-275217.7
         & 1.096 &                                           \\[1.5ex]
   \multirow{2}{*}{CI1042}   & 162527    &  0.119   & 2940728894
         & 0.406   & \multirow{2}{*}{0.006}    \\
                             & 162601    &  0.106   & 0084-01738
         & 0.400 &                                           \\[1.5ex]
   \multirow{2}{*}{CI0418}   & 184058    &  0.115   & 0139-01350                        & 0.2864   & \multirow{2}{*}{0.0105}    \\
      & 184031    &  0.642  & 26362                                & 0.2759  &                                           \\[1.5ex]
   \multirow{2}{*}{CI0961}   & 147928    &  0.094  & 2940682513  & 0.5963   & \multirow{2}{*}{0.0004}     \\
      & 147978    &  0.106   & 0082-00421                        & 0.5959   &                                          \\[1.5ex]
   \multirow{2}{*}{CI1905}   & 315366    &  0.106   & 0079-00072                        & 0.5815   & \multirow{2}{*}{0.2892}    \\
      & 315305    &  0.106   & 0078-01319                        & 0.2923   &                                         \\[1.5ex]
   \multirow{2}{*}{CI1906}   & 187271     &  0.189 & 34546                                      & 0.4357   & \multirow{2}{*}{0.2216}    \\
      & 187229     &  0.055 & J033244.87   & 0.2141   &                                         \\[1.5ex]
  \hline
  \hline
\end{tabular}}
\label{table:ir_cdfs_doubles_fusion_z}
\end{table}
\end{center}

\begin{center}
\tabcolsep=1pt
\begin{table}
 \centering
  \caption{Redshifts for possible IR doubles taken from OzDES, by a nearest-neighbour match between Fusion {\it Spitzer} and OzDES \citep{Yuan:2015} within $1''$. Of these 22, 20 have pairs of galaxies with $\Delta z/z_{\rm spec}<0.01$.}
{\renewcommand{\arraystretch}{1.1}%
  \begin{tabular}{@{}cccccc@{}}
   \hline
   \hline
   ATLAS  &   Fusion    & ang sep    & OzDES                               &  OzDES     &  OzDES                 \\
   ID     &   ID       & (arcsec)         &  ID                                    & z             & $\Delta z$                \\
   \hline
   \multirow{2}{*}{EI0151}       & 215007  & 0.224 & 2970674536                        & 0.12434   & \multirow{2}{*}{0.0004}  \\
         & 215061  & 0.191  & 2970674654  & 0.12478  &                                       \\[1.5ex]
   \multirow{2}{*}{EI0455}       & 221400  & 0.254 & 2971105989                        & 0.198   & \multirow{2}{*}{0.0036}  \\
         & 221459  & 0.136  & 0092-01998  & 0.195  &                                       \\[1.5ex]
   \multirow{2}{*}{EI0487}       & 101702  & 0.144 & J003459.03                        & 0.330   & \multirow{2}{*}{0.0011}  \\
         & 101761  & 0.111  & J003458.95  & 0.329  &                                       \\[1.5ex]
 \multirow{2}{*}{EI0863}       & 247565  & 0.100 & 0094-01686                       & 0.217   & \multirow{2}{*}{0.0064}  \\
         & 247574  & 0.333  & 2971175179  & 0.224  &                                       \\[1.5ex]
  \multirow{2}{*}{EI1034}       & 196663  & 0.20 & 0096-00993                        & 0.34711   & \multirow{2}{*}{0.0022}  \\
         & 196655  & 0.11  & 2971105849  & 0.3493  &                                       \\[1.5ex]
  \multirow{2}{*}{EI1219}       & 73440   & 0.21 & 2970777434    & 0.4001   & \multirow{2}{*}{0.0008}  \\
         & 73483   & 0.14 & 2970777513   & 0.3993  &              \\
  \hline
  \hline
\end{tabular}}
\label{table:ir_elais_doubles_fusion_z}
\end{table}
\end{center}



\subsection{Results of Cross-identification}

Table~\ref{table:results} presents the results of our cross-identification of the ATLAS catalogue with the Fusion catalogue. As we described in Section \ref{ozdes}, and illustrated in Figure \ref{fig:atlas_fusion_overlap}, $\sim96\%$ of total number of ATLAS sources are covered by the Fusion catalogue, which makes $2922$ radio sources in the CDFS field and $2113$ in the ELAIS-S1 field. So there are in total 5035 radio sources for XID with the Fusion catalogue. Not all radio sources we deal with have Fusion candidates inside of the search radius used in this work ($6''$). This number of ``blanks'' is small consisting of 222 for CDFS and 177 for ELAIS-S1. So the number of ``candidates'' (radio sources with one or more Fusion source(s) in the search area) drops to 2700 for the CDFS field and 1936 for the ELAIS-S1 field. We found that a large percentage of these candidate radio sources have just one Fusion source in the search radius $\sim60\%$  (see Table \ref{table:matches_distribution}). About $40\%$ of all (non-blank) radio sources have two or more (up to 5) Fusion sources within the search radius.

Applying the LR criteria for ''single'' sources and both LR and Reliability criteria for the situation when two or more Fusions sources are in the search radius, we find that about $\sim84\%$ of all candidates correspond to the criteria we use in this work for cross-identification in \S4.1 and \S4.2. The ATLAS sources without secure Fusion counterparts likely have counterparts below the Fusion detection limit.

\section{Host Galaxy Properties}

\begin{figure*}
    \centering
    \includegraphics[width=0.65\columnwidth]{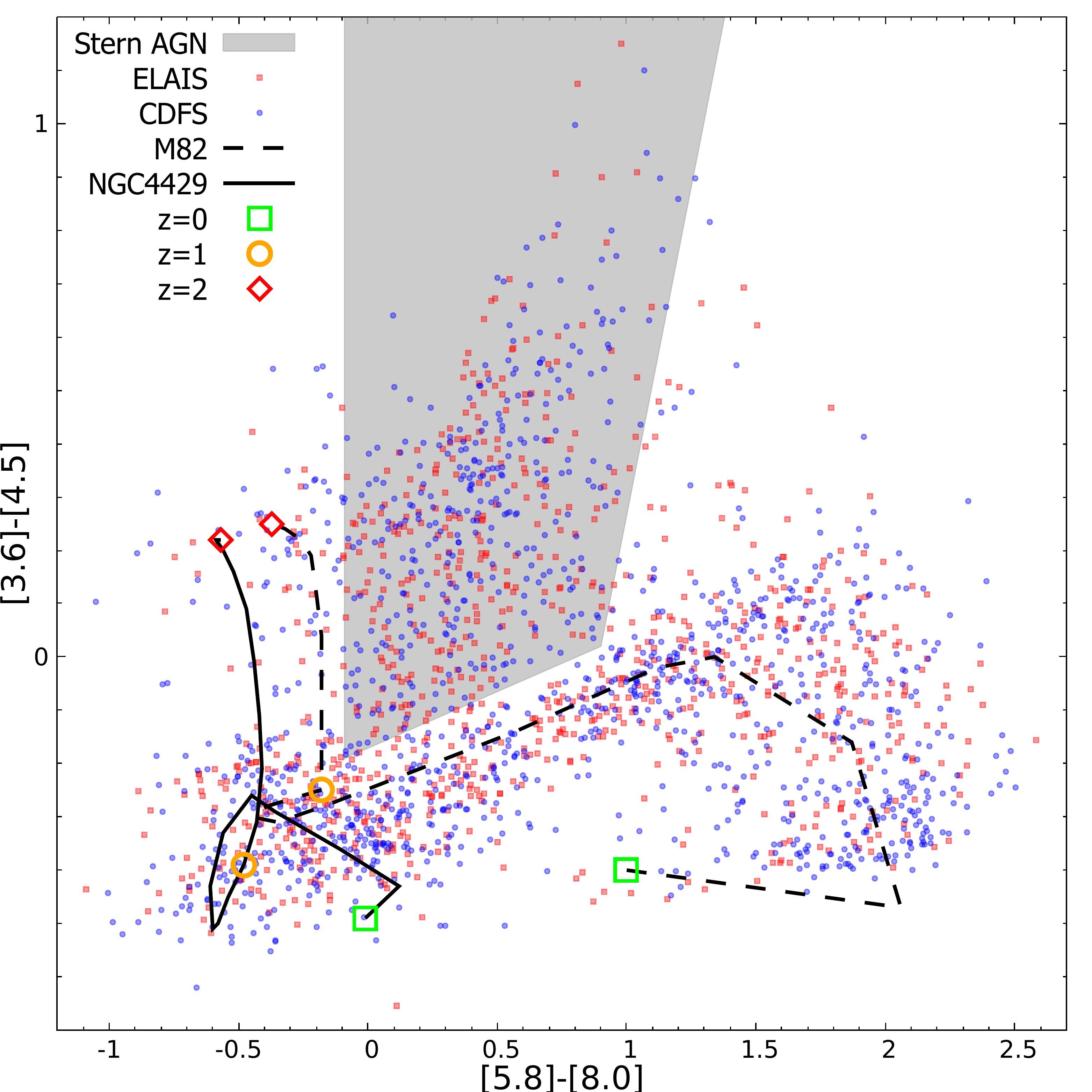}
    \includegraphics[width=0.65\columnwidth]{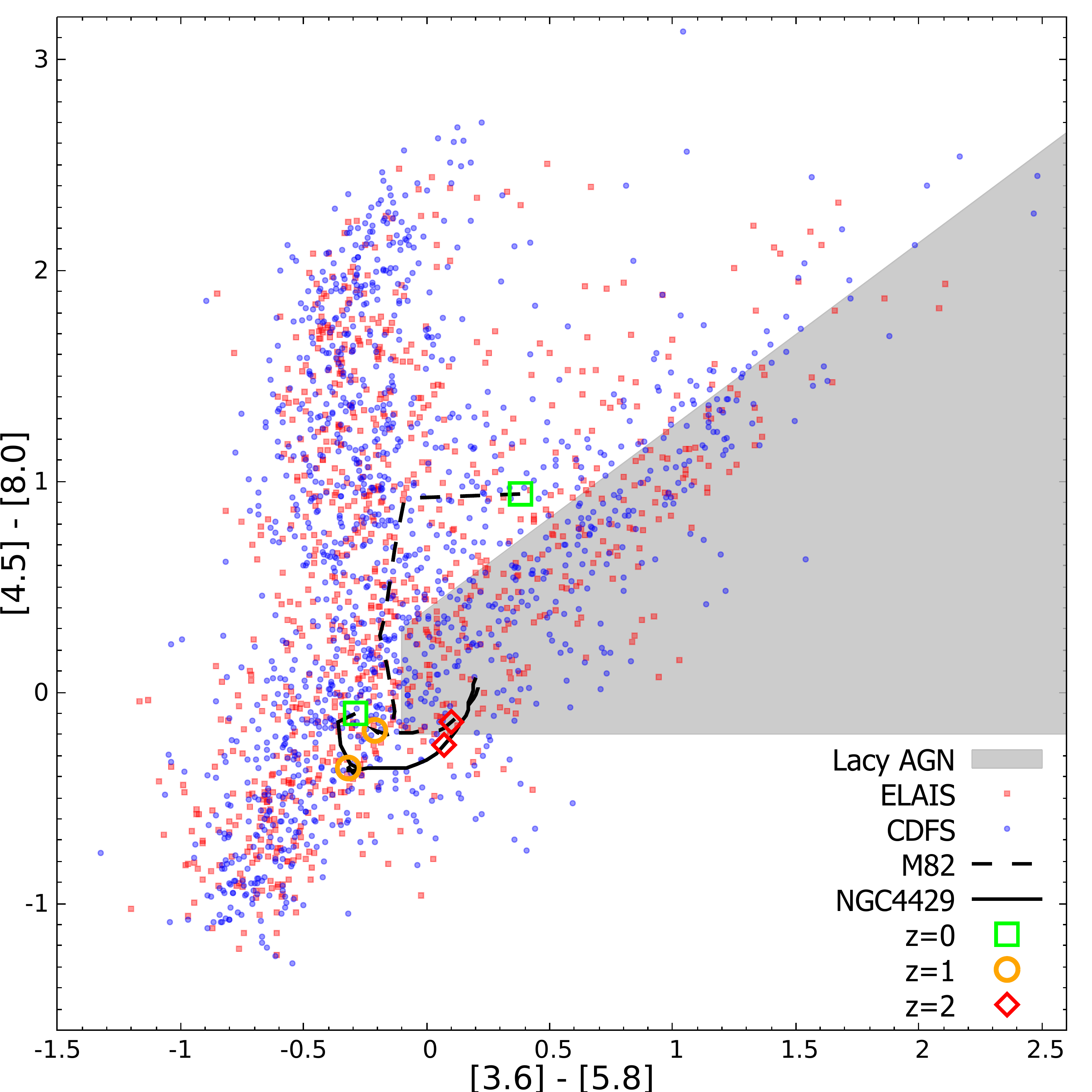}
    \captionof{figure}{We present the  colour-colour diagrams of the Fusion counterparts to the ATLAS sources (as determined in \S\ref{Rel_cutoff}). Top: is the $[5.8]-[8.0]$ vs. $[3.6] -[4.6]$ colour-colour plot. The grey shaded area shows the location of the \citet{Stern:05b} selection for AGN. Also included are the evolutionary tracks for M82 and NGC4429 from $z=0$ to $z=2$  taken from \citet{Seymour:2007}. Bottom: Using the same Fusion counterparts is the $[3.6] -[5.8]$ vs. $[4.5] -[8.0]$ colour-colour with the grey shaded area showing the location of the \citet{Lacy:04} selection for AGN. Again the evolutionary tracks for M82 and NGC4429 from $z=0$ to $z=2$  taken from \citet{Seymour:2007} are included.}
    \label{fig:colour-colour}
\end{figure*}

In this section we use the cross-identified sources in order to examine the nature of the faint ATLAS radio sources. The primary question we wish to address is whether our radio sources are AGN or SFGs. The advent of surveys with {\it Spitzer} has presented us with new methods of distinguishing between AGN and SFGs. These methods work by being sensitive to the hot ($\sim1000\,$K) dust around the AGN nucleus causing excess emission in the mid-IR compared to regular SFGs.

\subsection{IRAC Colour-Colour Plots}

Earlier work by 
\citet{Eisenhardt:2004} presented a vertical spur in the $[3.6]-[4.6]$ versus  $[5.8]-[8.0]$ colour-colour diagram which may be associated with AGN (where the magnitude difference $[i] - [j] = -2.5 \log\big(\frac{S_{i}}{S_{j}}\big)$, where $i$ and $j$ are the wavelengths of the {\it Spitzer} IRAC bands in $\mu$m). This is also supported by \citet{Stern:05b} who proposed a region in this parameter space which separates AGN from Galactic stars and SFG. \citet{Lacy:04} presented a $[8.0] - [4.5]$ versus $[5.8] - [3.6]$ colour-colour diagram and also identified an area to select AGN. 

As we now have cross-matches (based on the selection criteria in section \ref{Rel_cutoff}) of the ATLAS catalogue against the Fusion catalogue, we will use these to determine if a source is an AGN or SFG. 
We do so by plotting these sources in the same colour-colour diagrams as \citet{Stern:05b} and \citet{Lacy:04}. Not all of the cross-identifications  have infrared detections in the four {\it Spitzer} bands and, as such, comparisons can only be made where data is available at all wavelengths. We present the numbers of radio sources with cross-matches and their break down in Table~\ref{table:results2}.

Figure \ref{fig:colour-colour} contains two colour-colour plots for the Fusion counterparts to the ATLAS sources. Following \citet{Stern:05b}, the left plot presents the  $[3.6] -[4.5]$ versus $[5.8] -[8.0]$. The plot on the right is $[8.0] - [4.5]$ versus $[5.8] - [3.6]$ as per \citet{Lacy:04}. Evolutionary tracks from redshift 0 to 2 for a late type starburst galaxy (M82) and an early type galaxy (NGC 4429) are included in both plots. Markers have been placed to indicate $z = 0$, $z = 1$ and $z = 2$. This data has been taken from \citet{Seymour:2007} based on the work from \citet{Devriendt:1999}. We see how these evolutionary tracks generally remain outside the \citet{Stern:05b} and \citet{Lacy:04} AGN selection `wedge' (grey shaded areas) and neither would be selected as an AGN candidate if located below $z = 2$.

Many sources in the left hand plot of Figure \ref{fig:colour-colour} are spread along the evolutionary track of M82 to a redshift of $z=1$ as there are very few sources in the $z=1$ to $z=2$ region of this track. Of note is the vertical spur in the Stern AGN zone grey shaded consistent with \citet{Eisenhardt:2004} and \citet{Stern:05b}. In the right hand plot there is a clear fork with the right hand arm entering the Lacy AGN zone grey shaded. This is also consistent with \citet{Lacy:04}. We note that \citet{Mao2012}, using the ATLAS DR1 data release and associated spectroscopic classifications, showed that many spectroscopic AGN lay outside the Stern and Lacy wedges. 




By using the selection criteria in Section \ref{Rel_cutoff} for  Fusion cross-identifications, in the Lacy AGN selection zone there are a total of 848 XIDs and for the Stern AGN selection zone there are a total of 533 XIDs. A total of 956 XIDs satisfy the union of the Stern and Lacy selection criteria for AGN.


\subsection{Radio to $3.6\mu$m Flux Density Ratio}

\begin{figure}
    \includegraphics[width=\columnwidth]{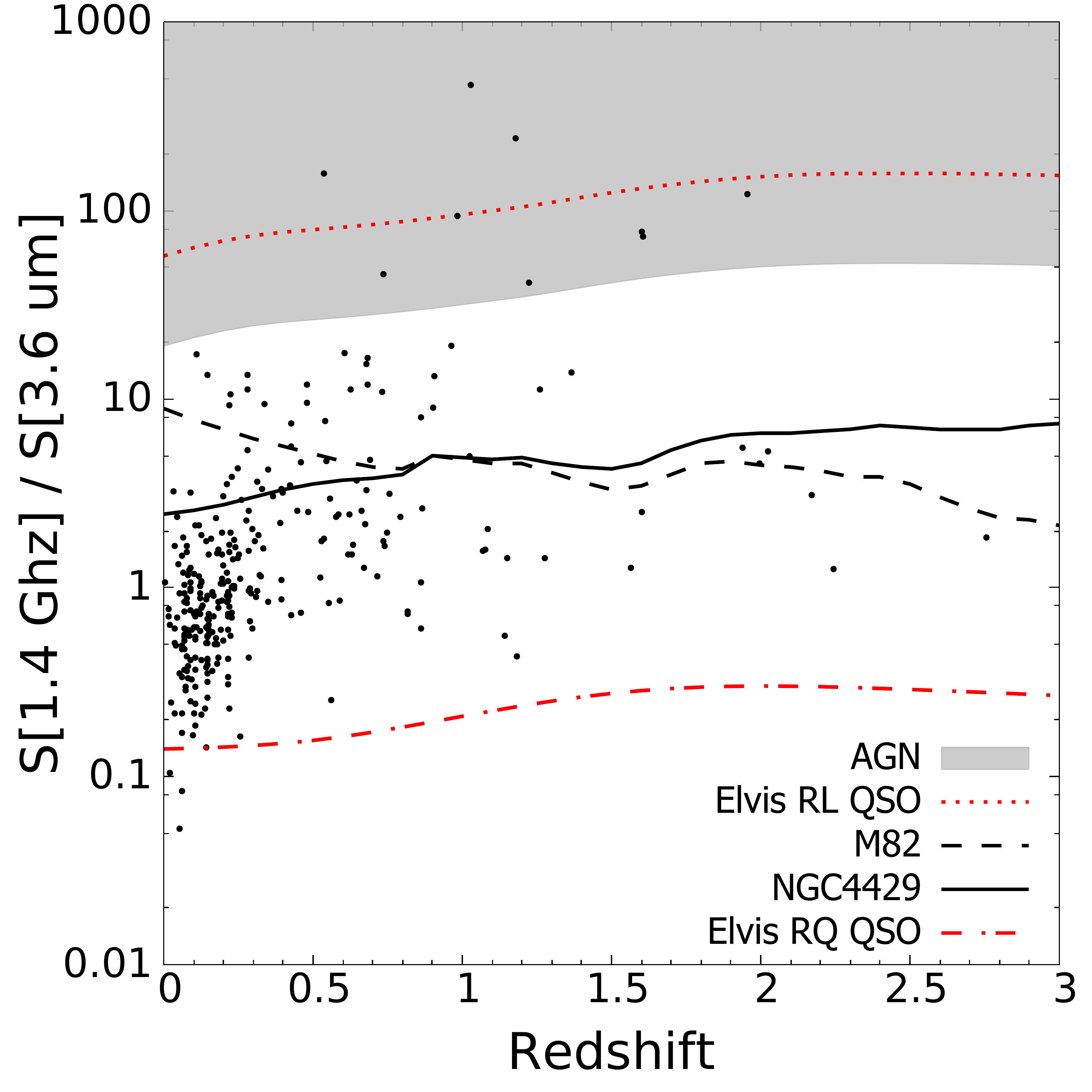}
    \captionof{figure}{The ratio between the radio $1.4\,$GHz and Fusion $3.6\,\mu$m flux density plotted as a function of redshift for all XIDs (determined in Section \ref{Rel_cutoff}). The red dotted line near the top of the figure indicates the loci of a classical radio-loud (radio-quiet) QSO \citet{Elvis:1994}, and the red dot-dashed line in the lower part of the figure indicates the loci for radio-quiet QSO \citet{Elvis:1994}. The grey shaded area denotes the population that we identify as the radio loud AGN. Also included are the evolutionary tracks for M82 and NGC4429 from $z=0$ to $z=3$  taken from \citet{Seymour:2007}. }
    \label{fig:radio_ir_flux_vs_redshift}
\end{figure}


In order to account for the relative radio emission from RLAGN, 
we examine the radio to $3.6\,\mu m$  flux density ratio of all the cross-matched sources with known redshifts in our sample in Figure \ref{fig:radio_ir_flux_vs_redshift}. We compare these to tracks of known sources shifted to higher redshifts (i.e. we compare the ratio of the observed frame $1.4\,$GHz and $3.6\,\mu$m flux densities shifted with redshift). For comparison we include the tracks for the radio-loud and radio-quiet AGN from \citet{Elvis:1994} and the two galaxies used in the previous section (the starburst M82 and the quiescent galaxy NGC4429). The various galaxy template tracks are relatively flat although some variation is seen with redshift. We note that the AGN templates are for unobscured AGN which do not include any potential obscuration of the AGN by dust from a torus or the host galaxy. Any obscuration would increase these flux ratios by suppressing the observed $3.6\,\mu$m flux density as it gets shifted to the near-infrared and optical rest-frame at higher redshift. 

The redshifts for 295 sources come from the OzDES global redshift catalogue \citep[][Lidman et al. in prep]{Yuan:2015}. We observe that most of our redshifts are at $z<0.3$, which is due to the targeting of the brightest optical counterparts by OzDES and earlier surveys \citep{Mao2012},
although there is a tail to $z\sim2.8$. This $z<0.3$ grouping typically have a flux ratio from $\sim 0.2$ to $2$, below that for the starburst and quiescent galaxy tracks. Why does our group of SFGs have lower flux ratios than these two sources? To first approximation we can say that, if these are star forming galaxies, the radio emission traces the star-formation rate (SFR) and the $3.6\,\mu m$  emission traces the stellar mass. Hence the galaxies in this grouping are likely similar to M82, but with a lower specific SFR (SFR per unit stellar mass). As the ATLAS sources are selected on a SFR proxy, radio flux, and lie at higher redshift, it is likely the higher stellar masses are pulling this observed ratio down despite the higher SFRs compared to M82. 

In terms of identifying which sources in this plot have radio emission powered by AGN, we can use the Elvis RL AGN track as a guide. Allowing for uncertainty in the models and the fluxes we suggest that any radio source with a ratio greater than one third of the track from the RL AGN, marked by the grey shaded area in Figure  \ref{fig:radio_ir_flux_vs_redshift}, is likely powered by an AGN. We find only nine sources above this line which also have a redshift value. We also have 57 matches with no redshift and the $3.6\,\mu m$ flux is below the detection limit. Taking their radio flux and dividing by the $3.6\,\mu m$ flux limit would place these matches in the grey AGN region. 


\subsection{Results of AGN Identification}


%

\begin{center}
\begin{table}
 \centering
  \caption{Results of the classification of cross-identification of ATLAS sources. The first row presents the total number of XIDs with a complete set of IRAC bands. The following rows show the AGN identified by the three methods (Stern, Lacy and Flux Density ratio) followed by the total number of AGN identified, i.e. the union of the preceeding three sets, and the percentage.}
  \begin{tabular}{@{}lccc@{}}
   \hline
   \hline
 Field & CDFS & ELAIS-S1 & both \\
    \hline
 \# with Fusion XID and  & 1153 & 834 & 1987\\
 \   all IRAC bands \\
 \# AGN & & & \\
 ~~~Stern & 298 & 235 & 533\\
 ~~~Lacy  & 490 & 358 & 848\\
 ~~~Ratio & 29 & 19 & 48\\
 Total $\cup$  & 550 (48\%) & 406 (49\%) & 956 (48\%)\\
  \hline
  \hline
\end{tabular}
\label{table:results2}
\end{table}
\end{center}

We have used the identification of Fusion counterparts to determine if our sample of ATLAS sources are AGN or SFG. We find 1987 cross matched radio sources with flux values for all four IRAC bands. Of these ($\approx27\,\%$) meet the Stern AGN selection criteria, and ($\approx43\,\%$) Lacy, in addition 48 ($\approx2\,\%$) of these sources lie above a line one third of the RL AGN track. Taking the union of all of these across our three selection criteria we find 956 ($\approx48\,\%$) are possible AGN.

\section{Catalogue and Code} \label{ATLAS_XID_Catalogue}

\subsection{Catalogue Description}


\begin{center}
\begin{table*}
\centering
\caption{ATLAS/FUSION SWIRE Cross-Identification Catalogue for the CDFS field. A description of the table is given in  Section \ref{ATLAS_XID_Catalogue}. (This table is available in its entirety in a machine-readable form in the online journal. A portion is shown here for guidance regarding its form and content, a full copy of the catalogue is available online.) }
\begin{tabular}{@{}lccccccccccc@{}}
\hline
\hline
(1) &  (2) & (3) & (4) & (5) & (6) & (7) & (8) & (9) & (10) & (11) \\
ATLAS &  $RA$      & $Dec$            &  $S_{\rm Int}$       & SWIRE  & $RA_{IR}$ & $Dec_{IR}$ & $S_{3.6\mu{\rm m}}$ & $\sigma_{3.6\mu{\rm m}}$ & $\log_{10}(LR)$ & Reliability  \\
ID       &   (J2000)       & (J2000)         & mJy           & ID          &  deg             &          deg       & $\mu$Jy                      &  $\mu$Jy             &       &       & \\
\hline
CI0001C1 & 52.1516 & -28.6982 & 132.5 & 432065 & 52.1520 & -28.6988 & 2565 & 7.07 & -1.154 & 0.290 \\
CI0002 & 51.8917 & -28.7726 & 158.151 & 428929 & 51.8917 & -28.7725 & 101.84 & 0.8 & 2.556 & 0.999 \\
CI0003 & 53.5387 & -28.4055 & 74.8 & 158805 & 53.5388 & -28.4053 & 67.29 & 1.09 & 2.011 & 0.998 \\
CI0005C1 & 51.9088 & -28.0239 & 19.8 & 456752 & 51.9079 & -28.0232 & 5.16 & 0.43 & -6.966 & 0.000 \\
CI0005C2 & 51.9127 & -28.0357 & 0.692 & 456300 & 51.9127 & -28.0358 & 24.75 & 0.81 & 1.906 & 0.997 \\
CI0005C3 & 51.9067 & -28.0251 & 69.6 & 456683 & 51.9071 & -28.0250 & 200.42 & 1.63 & 1.424 & 0.993 \\
CI0007 & 53.9722 & -27.4613 & 118.207 & 63449 & 53.9722 & -27.4612 & 398.33 & 2.01 & 2.458 & 0.999 \\
CI0008 & 52.1943 & -28.4379 & 56.6902 & 303864 & 52.1940 & -28.4383 & 68.82 & 0.88 & 0.693 & 0.966 \\
CI0009 & 53.8646 & -27.3308 & 95.6314 & 209688 & 53.8646 & -27.3307 & 157.62 & 1.77 & 2.610 & 0.999 \\
CI0010 & 53.6121 & -27.7338 & 55.7524 & 190007 & 53.6121 & -27.7338 & 201.51 & 1.55 & 2.546 & 0.999 \\
\hline
\hline
\end{tabular}
\label{table:CDFS_XID_CATALOGUE}

\bigskip


\caption{ATLAS/FUSION SWIRE Cross-Identification Catalogue for the ELAIS-S1 field. A description of the table is given in Section \ref{ATLAS_XID_Catalogue}. (This table is available in its entirety in a machine-readable form in the online journal. A portion is shown here for guidance regarding its form and content, a full copy of the catalogue is available online.)  }
\begin{tabular}{@{}lccccccccccc@{}}
\hline
\hline
(1) &  (2) & (3) & (4) & (5) & (6) & (7) & (8) & (9) & (10) & (11) \\
ATLAS &  $RA$      & $Dec$            &  $S_{\rm Int}$       & SWIRE  & $RA_{IR}$ & $Dec_{IR}$ & $S_{3.6\mu{\rm m}}$ & $\sigma_{3.6\mu{\rm m}}$ & $\log_{10}(LR)$ & Reliability  \\
ID       &   (J2000)       & (J2000)         & mJy           & ID          &  deg             &          deg       & $\mu$Jy                      &  $\mu$Jy             &       &       & \\
\hline
EI0001 & 9.06966 & -43.15964 & 160.032 & 87322 & 9.06957 & -43.15968 & 103.24 & 1.53 & 2.565	 & 0.999 \\
EI0002C2 & 7.87001 & -43.68909 & 193.85 & 373161 & 7.87013 & -43.68917 & 46.44 & 0.96 & 2.038 &	0.998 \\
EI0003 & 8.28863 & -43.99076 & 69.6057 & 220919 & 8.28844 & -43.99079 & 141.9 & 1.8 & 2.554	& 0.999 \\
EI0004C1 & 8.67872 & -43.50959 & 49.97 & 237221 & 8.67851 & -43.50948 & 89.18 & 1.45 & 2.352	 & 0.999 \\
EI0004C3 & 8.67261 & -43.51230 & 0.538 & 237403 & 8.67237 & -43.51141 & 228.49 & 2.27 & -3.161 &	0.003 \\
EI0005 & 8.01844 & -44.19189 & 35.5887 & 350583 & 8.01828 & -44.19192 & 48.61 & 0.84 & 2.042	 & 0.998 \\
EI0006 & 8.20550 & -44.36404 & 39.7013 & 341374 & 8.20543 & -44.36403 & 373.25 & 2.28 & 2.420 &	0.999 \\
EI0007 & 9.34721 & -44.37919 & 44.6797 & 159474 & 9.34722 & -44.37909 & 51.05 & 0.81 & 2.308	 & 0.999 \\
EI0008 & 9.19112 & -43.09654 & 30.4775 & 87851 & 9.19100 & -43.09654 & 10.76 & 0.68 & 1.705 &	0.996 \\
EI0009C1 & 9.32550 & -44.50327 & 49.35 & 162876 & 9.32449 & -44.50391 & 133.17 & 1.3 & -7.356 &	0.000 \\
\hline
\hline
\end{tabular}
\label{table:ELAIS_XID_CATALOGUE}

\end{table*}
\end{center}

In this section we describe the catalogue containing the results of the ATLAS cross-identification with Fusion using the LRPY algorithm discussed in this paper. The information is divided into two tables, one for each field CDFS and ELAIS-S1. Example subsets are given in Table \ref{table:CDFS_XID_CATALOGUE} for CDFS and Table \ref{table:ELAIS_XID_CATALOGUE} for ELAIS-S1. The cross identification catalogue columns are organized as follows:

\begin{itemize}[leftmargin=1cm]
      \item[] \textit{Column (1)}   - ATLAS DR3 Identification number of the radio source ``cid''
      \item[] \textit{Column (2)}   - Right Ascension (J2000) of the radio source, decimal degrees
      \item[] \textit{Column (3)}   - Declination (J2000) of the radio source, decimal degrees
      \item[] \textit{Column (4)}   - Integrated Radio Flux Densities ($\mu$Jy) at 1.4\,GHz
      \item[] \textit{Column (5)}   - Fusion Identification number ``swire\textunderscore index\textunderscore {\it Spitzer}''	
      \item[] \textit{Column (6)}   - Right Ascension (J2000) of the IR candidate, decimal degrees
      \item[] \textit{Column (7)} - Declination (J2000) of the IR candidate, decimal degrees
      \item[] \textit{Column (8)}   - IR Flux Density at $3.6\,\mu$m ($\mu$Jy)
      \item[] \textit{Column (9)}   - IR Flux Density Uncertainty at $3.6\,\mu$m ($\mu$Jy)
      \item[] \textit{Column (10)} - $Log_{10}$ of Likelihood Ratio of the IR candidate
      \item[] \textit{Column (11)} - Reliability of the IR candidate
\end{itemize}

These two tables are available in their entirety including all Fusion sources within $6''$ of an ATLAS source in a machine-readable format in the supplementary material. No filtering on Reliability or Likelihood Ratio has been undertaken. Column 10 has been presented with its $Log_{10}$ value to make the column width manegable, also Column 11 is presented to three decimal places. Both columns in the supplementary material will be presented to a higher precision.  From our findings in Section \ref{Rel_cutoff} we recommend the following selection criteria for accepting identifications: $L_{c} \geq 0.01$ and the Reliability is greater than $R$ limit given by using Equation~\ref{eq:rel_sym} with $\beta = 0.4$.
The data is also available as a series of normalised relational database tables where, by using the index columns ``cid'' and ``swire\textunderscore index\textunderscore spitzer'' as the relationship to join the tables, it is possible for the reader to construct their own version of the catalogue or work with the data in other ways. 


%

%
%


\subsection{The LRPY Algorithm}

The method described in Section \ref{scmt} along with the selection rules presented in Section \ref{Analysis} have been coded in Python and the full set of files is available from github\footnote{https://github.com/sdweston/LikelihoodRatio} under a GNU General Public License (GPL) V3.0. We intend to use this code on the ASKAP/EMU survey as one of several complimentary methods to determine counterparts to the estimated $7\times10^7$ faint radio sources. Anyone is permitted to use this code for research purposes and we ask that they cite this paper.

\section{Conclusions}


We have developed a Python based code to allow cross-matching of lower resolution survey data to higher source density, higher resolution catalogues implementing the Likelihood Ratio technique. Our motivation is to use this code as one of several techniques to cross-match the ASKAP/EMU survey to higher resolution optical/near-IR data. This code, LRPY, accurately accounts for sources below the detection limit in an automated fashion. It is suitable for any cross-matching of catalogues with significantly mismatched resolutions. 
We have added an extension to the LR algorithm to identify potential multiple matches to a lower resolution source, i.e. doubles.  We make this code publicly available to the community. Future updates to this code will include the (optional) ability to handle more complex radio sources (e.g. radio doubles and triples). 

We have used LRPY to cross-match the ATLAS DR3 radio survey to the {\it Spitzer} Data Fusion catalogues with a search radius of $6''$.
Setting the possible false detection rate of $5\,\%$, and using the new resultant likelihood ratio and Reliability cutoff selection criteria described in Section 4.1, rather than simply a reliability threshold, we obtain $2222$ ($82\%$) matches in the CDFS field and $1626$ ($83\%$) in the ELAIS-S1 field. 
 Of these matches, we obtain a subset with detections in all four IRAC bands consisting of $2133$ sources ($1243$ for CDFS and $890$ for ELAIS-S1). A much smaller subset has redshifts consisting of $295$ sources ($186$ CDFS and $109$ for ELAIS-S1). 
Hence, from this work we present a new catalogue listing ATLAS DR3 radio sources with their {\it Spitzer} Data Fusion counterparts including the likelihood and reliability figure to allow the reader to use their own selection criteria as required.

We have identified a subset of $64$ {\it Spitzer} Data Fusion doubles ($38$ in CDFS and $26$ in ELAIS-S1), i.e. radio sources with two Spitzer Data Fusion candidates meeting our selection criteria in Section \ref{ir_doubles}. From these pairs we find 22 with a redshift for each member; we find 20 of these have a $\Delta z/z < 0.01$ and we identify them as potentially interacting galaxies contributing to the one radio source. Two pairs are confirmed as interacting galaxies from deep {\it HST} imaging.


Taking the available {\it Spitzer} Data Fusion colour-colour information for the possible matches we present their characteristics with respect to the Stern and Lacy AGN selection criteria. For the two fields we identify 848 AGN radio sources using the Lacy selection criteria which is $\approx42\,\%$ of the candidates, and 533 if using the Stern criteria which is $\approx27\,\%$. Also, we examine the radio to $3.6\,\mu$m flux density ratio as a function of redshift to search for radio-loud AGN. We find a cluster of objects at $z<0.3$ and flux ratio $~0.2$ to $2$ which we surmise are SFG, but with a lower relative SFR to stellar mass than the modeled M82 track. We propose a cutoff where the flux ratio is greater than one third of the value of the RL-AGN track. Taking the union of all three AGN selection criteria we identify 956 $\approx48\,\%$ possible AGN. 



\section*{Acknowledgements}

We thank L. Dunne and S. Maddox for hosting the author and for initial advice at the University of Canterbury and subsequent helpful discussions. 
Stuart Weston is a recipient of a CSIRO Astronomy \& Space Science, Australia Telescope National Facility Graduate Research Scholarship. Nicholas Seymour is the recipient of an Australian Research Council Future Fellowship. Mattia Vaccari acknowledges support from the European Commission Research Executive Agency (FP7-SPACE-2013-1 GA 607254), the South African Department of Science and Technology (DST/CON 0134/2014) and the Italian Ministry for Foreign Affairs and International Cooperation (PGR GA ZA14GR02). We thank the referee, Jim Condon, for critically reading the manuscript and suggesting substantial improvements which have significantly improved the paper.

\clearpage
\bibliographystyle{mnras}
\bibliography{mnemonic,biblio}

\clearpage
\appendix
\section{Postage Stamp Images of XID Examples}\label{App:Appendix_A}

\FloatBarrier

\noindent\begin{minipage}{\textwidth}
    \captionsetup{type=figure}
    \centering
    \includegraphics[width=0.49\textwidth]{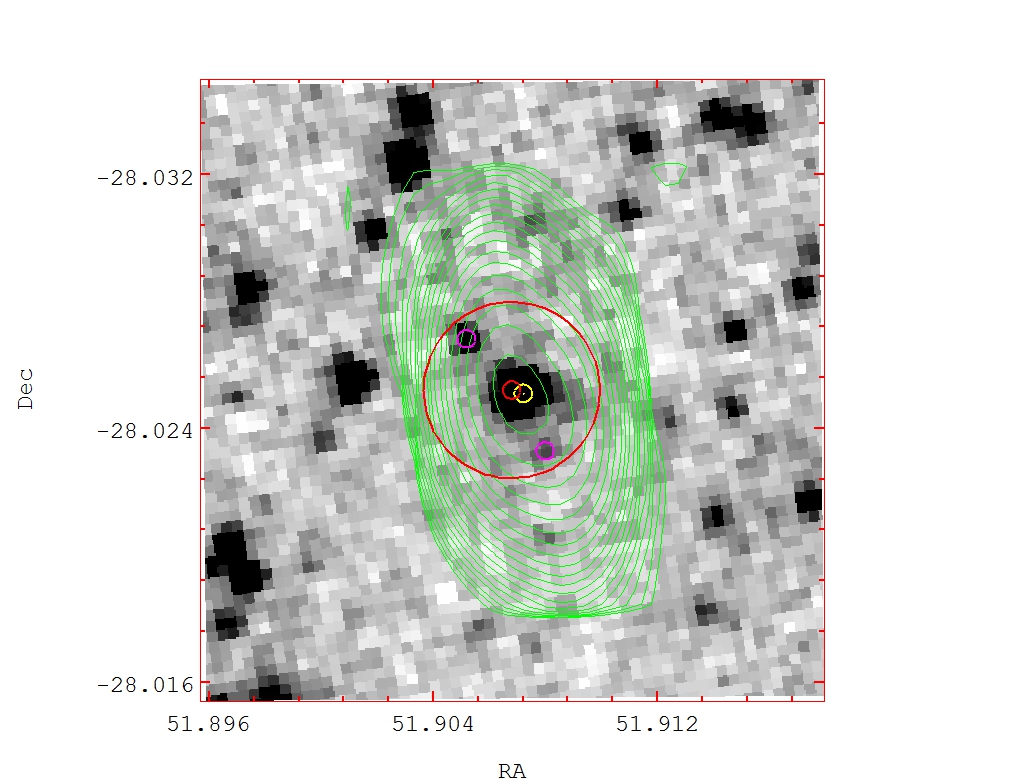}
    \includegraphics[width=0.49\textwidth]{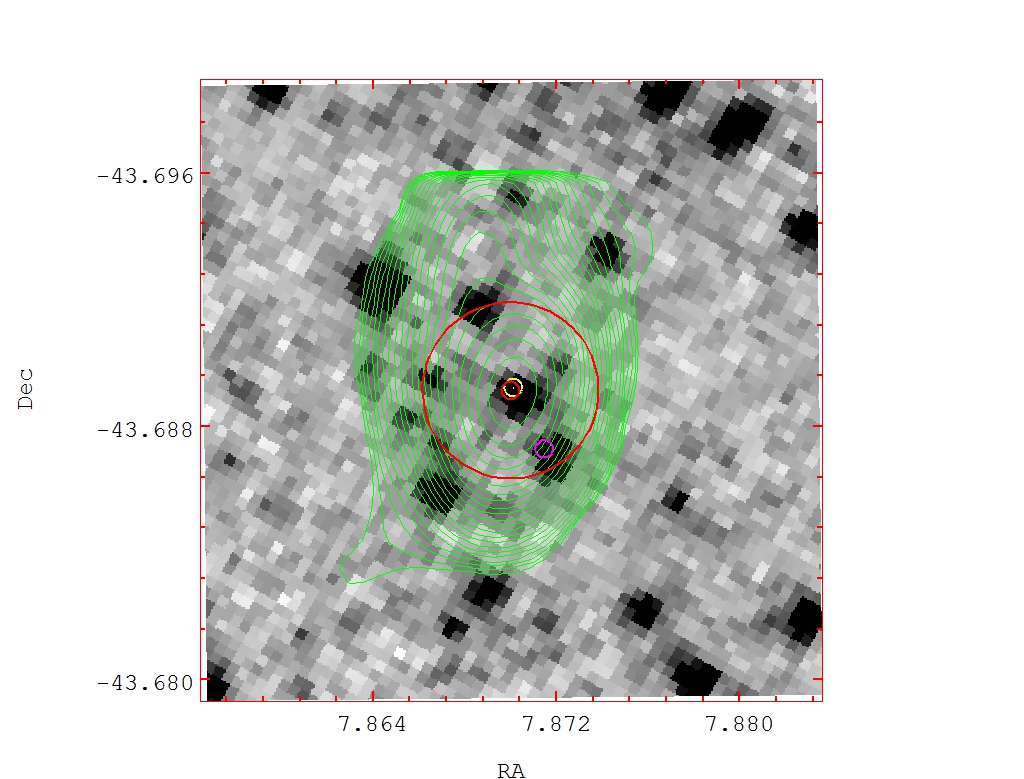}
\caption{Example postage stamps of the cross-identifications of ATLAS sources with the Fusion catalogue. We overlay ATLAS contours (starting at $3\sigma$ and then spaced by a factor of 2) on greyscale $3.6\,\mu$m images centred on the radio source position  to visually demonstrate the LRPY XID. The small red circle denotes the ATLAS radio candidate position and the larger red circle is the $6''$ search radius. The small yellow circle denotes the candidate Fusion counterpart position with the selection criteria give in \S4.1; and the small magenta circles are other candidate Fusion counterparts within the search radius that have a Reliability and LR outside the selection criteria. Each image is $70''$ square. On the left is ATLAS source CI0005C3 and on the right is ATLAS source EI0002C2}
\label{fig:AppendixXIDPS}
\end{minipage}

\noindent\begin{minipage}{\textwidth}
    \captionsetup{type=figure}
    \centering
    \includegraphics[width=0.49\textwidth]{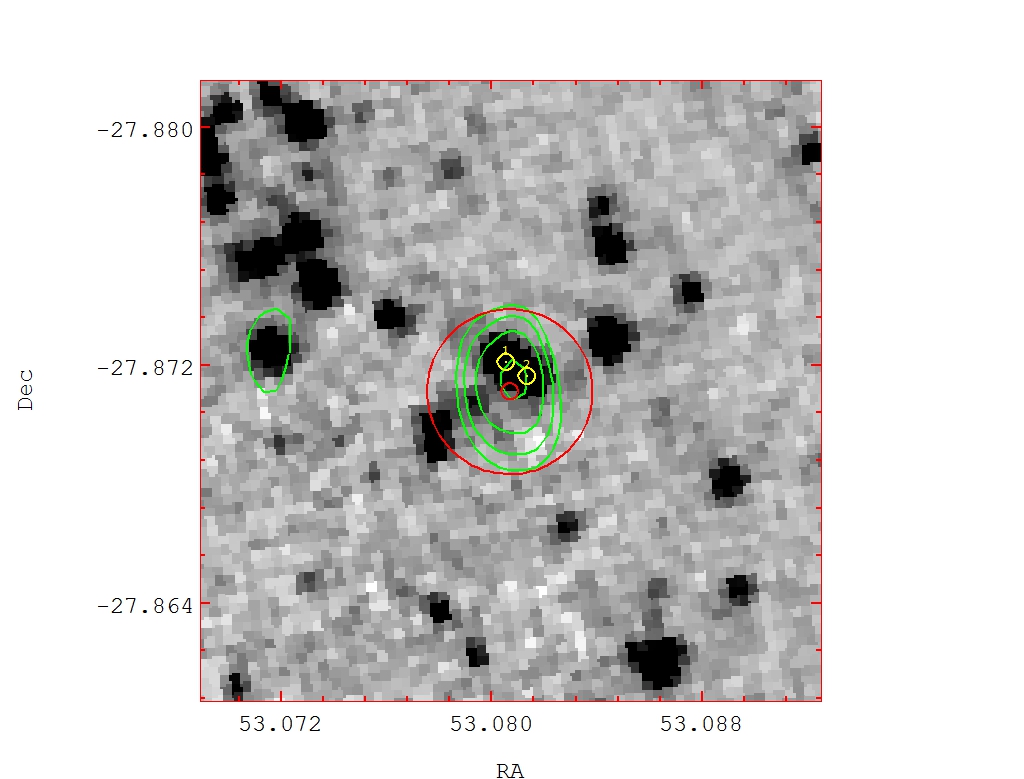}
    \includegraphics[width=0.49\textwidth]{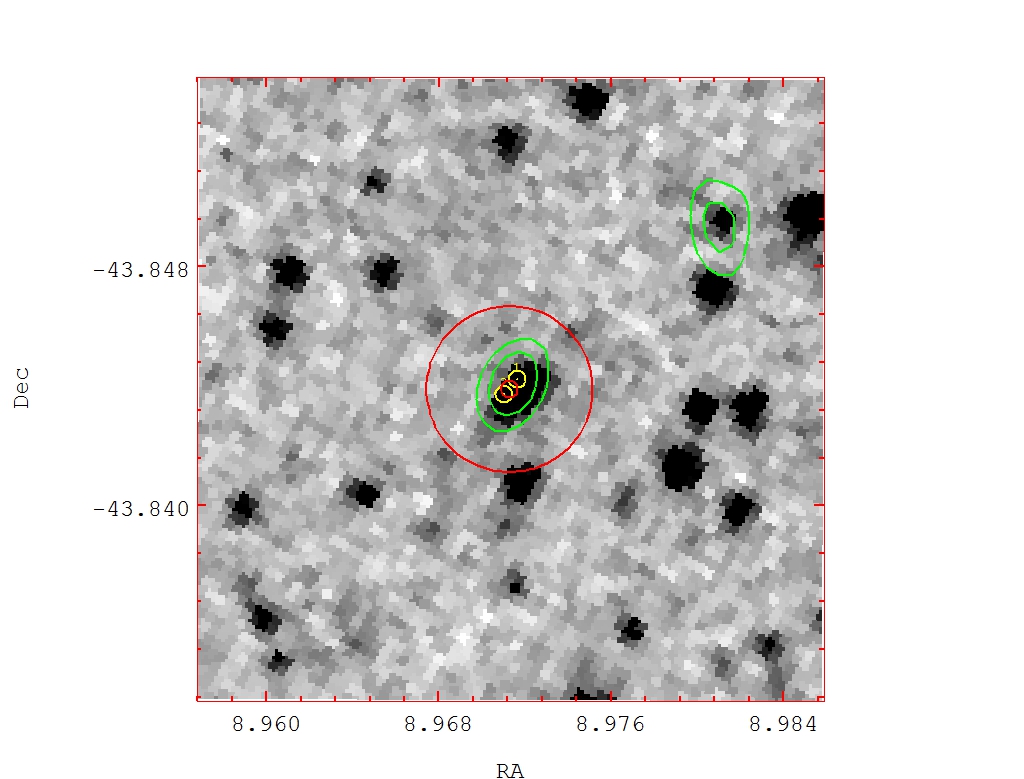}
    \caption{Example postage stamps of Infared Double (IRD) candidates with the radio contours (starting at $3\sigma$ and then spaced by a factor of 2) overlaid on greyscale $3.6\,\mu$m images centred on an ATLAS radio source position. The small red circle denotes the ATLAS radio candidate position and the larger red circle is the $6''$ search radius. The small yellow circles denote the possible IRD candidate positions using the selection criteria in Section \ref{ir_doubles}. Each image is $75''$ square. On the left is ATLAS source CI1036 and on the right is ATLAS source EI1034 with the two IR candidates which are in Table \ref{table:ir_doubles_fusion_z} with similar $z$.}
\label{fig:AppendixIRDoublesPS}
\end{minipage}

\noindent\begin{minipage}{\textwidth}
    \captionsetup{type=figure}
    \centering
    \includegraphics[width=0.49\textwidth]{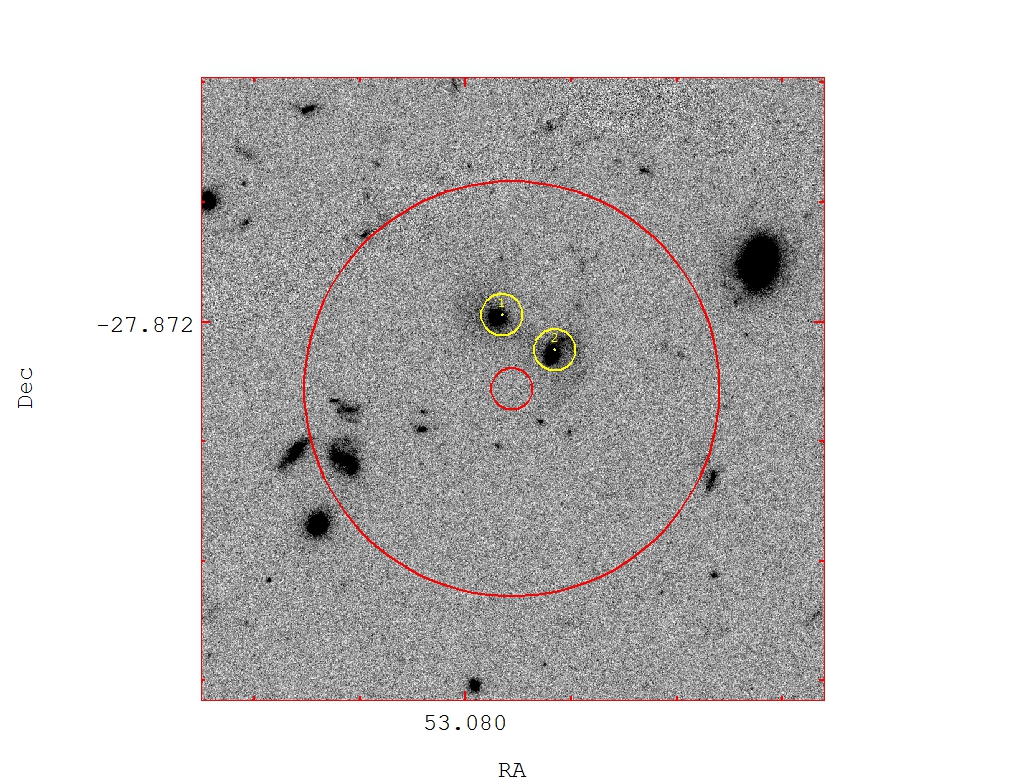}
    \includegraphics[width=0.49\textwidth]{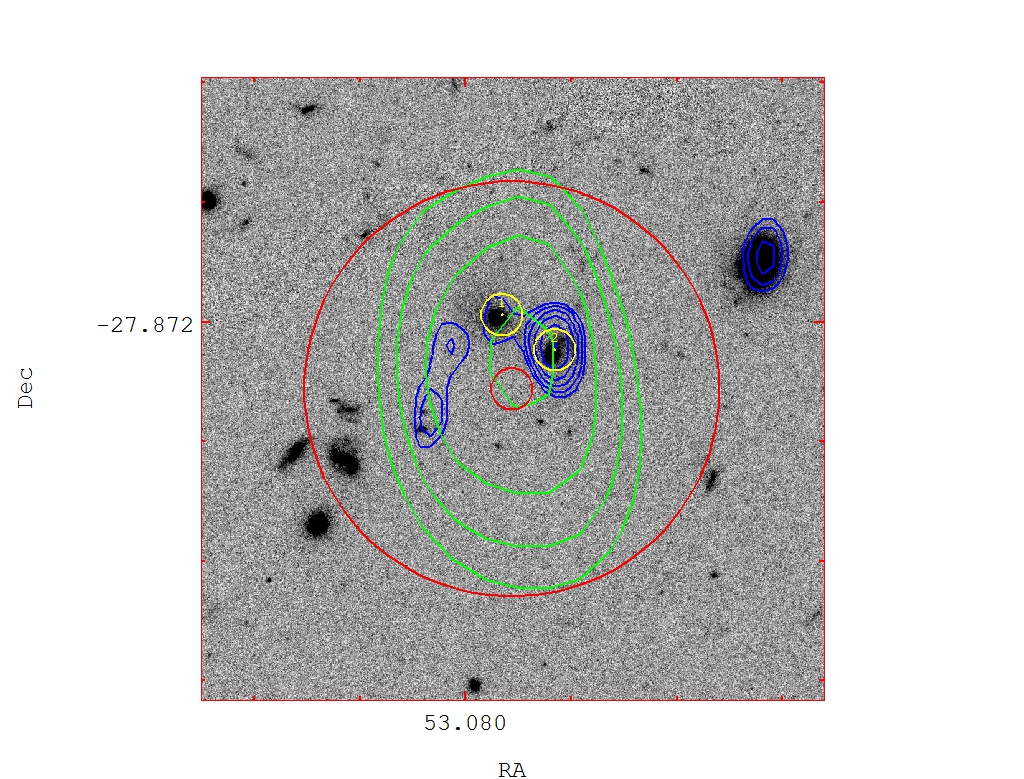}
    \caption{Postage stamp HST image centered on the ATLAS source CI1036 position, each image is $30''$ square. On the left hand image the Infared Double (IRD) candidates are marked with open yellow circles, and the ATLAS radio position with a small red open circle, the larger open red circle is the $6''$ search radius. The IR Source marked 1 is $z=1.0967$ and source 2 is $z=1.0960$ where $\delta z = 0.0007$. The right hand image has the ATLAS radio contours (starting at $3\sigma$ and then spaced by a factor of 2) overlaid in green, in addition the VLA radio contours in blue (using the same contour spacing as for ATLAS).}
\label{fig:AppendixIRD_CI1036}
\end{minipage}

\noindent\begin{minipage}{\textwidth}
    \captionsetup{type=figure}
    \centering
    \includegraphics[width=0.49\textwidth]{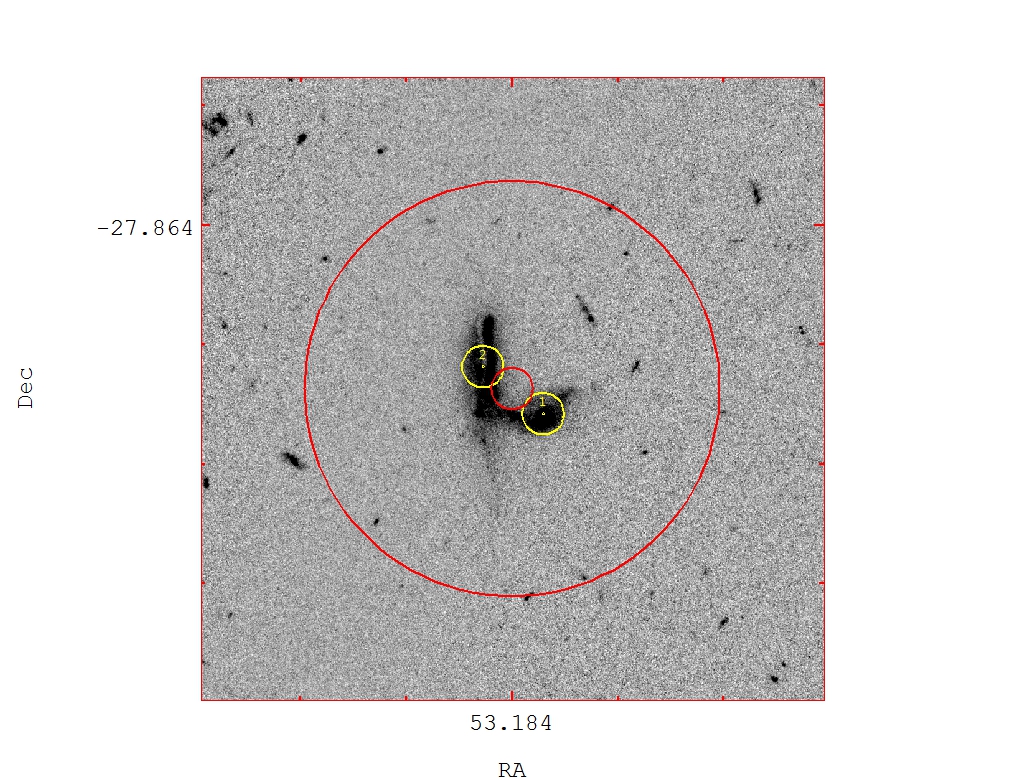}
    \includegraphics[width=0.49\textwidth]{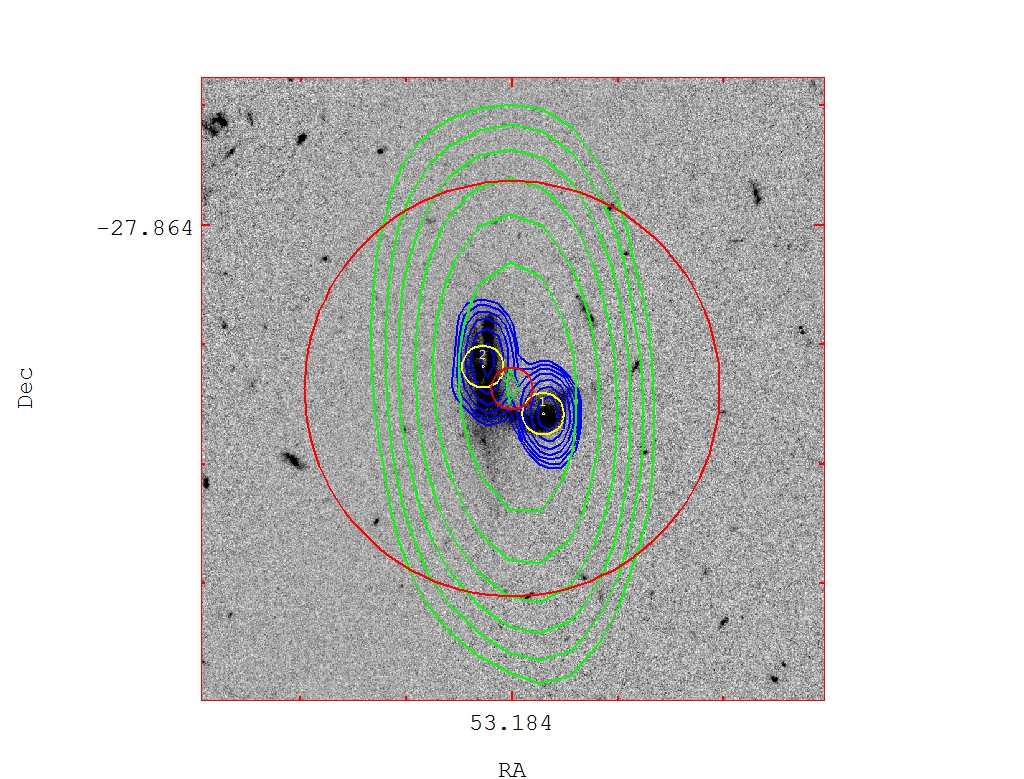}
    \caption{Postage stamp HST image centred on the ATLAS source CI0418 position; each image is $30''$ square. On the left hand image the Infared Double (IRD) candidates are marked with open yellow circles, and the ATLAS radio position with a small red open circle; the larger open red circle is the $6''$ search radius. The IR Source marked 1 is $z=0.2864$ and source 2 is $z=0.2759$ where $\delta z = 0.0105$. The right hand image has the ATLAS radio contours (starting at $3\sigma$ and then spaced by a factor of 2) overlaid in green; in addition the VLA radio contours in blue (using the same contour spacing as for ATLAS).}
\label{fig:AppendixIRD_CI0418}
\end{minipage}


\end{document}